\documentclass{sapm}
\usepackage{geometry}                		
\usepackage{graphicx}				
\usepackage{amssymb,amsmath,amsthm}
\usepackage{subcaption}

\jname{STUDIES IN APPLIED MATHEMATICS}
\jvol{AA}
\jyear{YYYY}
\doi{10.1111/((please add article doi))}

\begin{document}

\title{Elastic sheets, phase surfaces and pattern universes}
\author[A.C.Newell  and S.C.Venkataramani]{Alan C. Newell \thanks{Dave Benney made a huge difference in so many lives, and especially in those of his students. He rescued me (ACN) when I was on the verge of flunking out. He listened, he encouraged, he rarely disparaged, he coaxed, he had limitless patience and got the best out of each and every one of us. His door was always open. He had a wonderful sense of low key humor although some of his wittiest remarks could take a while to sink in. His pioneering scientific contributions include the role of three dimensionality in the onset of turbulence, waveturbulence closures, multiphase waves and nonlinear critical layers. He was a first rate department chairman. With his passing, the mathematical sciences lost one of their very best. His like does not often come along.} and Shankar C. Venkataramani \thanks{email for correspondence: shankar@math.arizona.edu}} \affil{Dept. of Mathematics, University of Arizona, Tucson, AZ 85721} \affil{Dept. of Mathematics, University of Arizona, Tucson, AZ 85721}

\date{}							

\maketitle

\begin{abstract}
We  connect the theories of the deformation of elastic surfaces and phase surfaces arising in the description of almost periodic patterns. In particular, we show parallels between asymptotic expansions for the energy of elastic surfaces in powers of the thickness $h$ and the free energy for almost periodic patterns expanded in powers of $\epsilon$, the inverse aspect ratio of the pattern field.  For sheets as well as patterns, the resulting energy can be expressed in terms of natural geometric invariants, the first and second fundamental forms of  the elastic surface, respectively the phase surface. We discuss various results for these energies and also address some of the outstanding questions. 

We extend previous work on point (in 2D) and loop (in 3D) disclinations and connect their topological indices with the condensation of Gaussian curvature of the phase surface. Motivated by this connection with the charge and spin of pattern quarks and leptons, we lay out an ambitious program to build a multi-scale universe inspired by patterns in which the short (spatial and temporal) scales are given by a nearly periodic microstructure and whose macroscopic/slowly varying/averaged behaviors lead to a hierarchy of structures and features on much longer scales including analogues to quarks and leptons, dark matter, dark energy and inflationary cosmology. One of our new findings is an interpretation of dark matter as the energy density in a pattern field. The associated gravitational forces naturally result in galactic rotation curves that are consistent with observations, while simultaneously avoiding some of the small-scale difficulties of the standard $\Lambda$CDM paradigm in cosmology.
\end{abstract}


\section{Introduction} \label{sec:intro}

Patterns of an almost periodic nature are ubiquitous. One sees them as ripples on long sandy beaches, in cloud structures, in geological formations such as the Giant's causeway, on the shoot apical meristems of plants and flowers, on animal coats and fish skins, on the tips of fingers, on palms and soles. They are also observed in laboratory experiments on convection, liquid crystals, the wrinkling and buckling of elastic sheets and shells, colliding light beams, lasers, and indeed in many condensed matter contexts. They arise in systems driven far from equilibrium by some external stress which, when exceeding a critical threshold, causes a uniform state with many symmetries, for example continuous translations and rotations, to become unstable, and some but not all of the symmetries are broken. In the wake of the instability, various shapes and configurations with less symmetry are preferentially amplified. They compete for dominance and eventually a winning combination emerges. The winning configuration often exhibits a preferred wavelength. In planar geometries, most often, the local winning planform consists of a set of rolls (or stripes) because the nonlinear cross coupling in multi-mode arrangements tends to raise the energy. One exception to this general observation is the multi-mode arrangement leading to hexagonal structures in planar geometries with rotational symmetry. A hexagonal structure can reflect an in-out or up-down asymmetry in the radial and vertical directions as,  for example, in spherical shells under radial compression and convection layers heated from below. The presence of such an asymmetry is captured by  quadratic nonlinearities in the evolution of the pattern thereby breaking the $w \to - w$ invariance. 
  
 Determining the preferred local waveform is not the end of the story in describing such patterns. The orientations of these winning local planforms are not chosen by energetic or dynamical considerations. Instead, they are chosen by local biases, reflecting the fact that the system is rotationally invariant. For example, in a convection layer with a heated horizontal boundary, the local rolls will have wavevectors parallel to the normal to the boundary. The epidermal ridges on fingertips align themselves locally with nail boundaries and crease furrows. As a result, natural patterns consist of a mosaic of patches of the preferred planform with different orientations, separated one from another by line and point defects.
  
It is this class of systems, namely a system described by a real microscopic field $w$, whose uniform state enjoys a continuous rotational and translational symmetry, the instability of which involves a preferred length scale $2 \pi/k_0$, which is the focus of this paper. Far from the onset of this instability, the microscopic field can be approximately described by it local periodic structure
\begin{align}
\label{local-periodic}
w & = f(\theta;\{A_n\},R) + \mbox{ corrections of order } \epsilon = \frac{\lambda}{L} \\
f(\theta) & = \sum A_n(k,R) \cos(n \theta) \nonumber
\end{align}
where $f$ is $2 \pi$ periodic in the phase $\theta$, $\mathbf{k} = \nabla \theta$ is slowly varying in space and time, and the sequence $\{A_n\}$ consists of slaved amplitudes. The symmetry $\theta \to \theta + \theta_0$ for arbitrary $\theta_0$ means that the determination of the corrections in \eqref{local-periodic} demand a compatibility condition (cf. Fredholm alternative) which constrains how the phase $\theta$ and its gradient $\mathbf{k} = \nabla \theta$ evolve. The evolution equation is a gradient flow for an appropriate energy functional. The equation has a universal form and is known as the Cross-Newell equation \cite{Cross_Convection_1984,Passot_The_1991}. An outline of its derivation is given in section~\ref{sec:patterns}. What is of most interest to us in this paper, however, is its energy functional, which we derive directly by averaging the energy functional of an underlying microscopic system which we assume to be gradient. The energy functional for patterns  takes a form which is very similar to that of the energy functional for an elastic sheet in that it consists of two terms at order $\epsilon$ and $\epsilon^3$ which are close analogues of the strain and bending energies of the elastic sheet. For patterns, the analogous strain energy is measured by the integrated root-mean-square deviation of the local pattern wavenumber from its preferred value. It has been known for some time \cite{Love_Treatise_Elasticity,Fox_A_1993} that, under circumstances we outline in section~\ref{sec:elastic}, the elastic sheet energy can be written in terms of geometrical quantities, namely invariant combinations of the elements of the metric and curvature tensors associated with the deformed surface. We show in section~\ref{sec:patterns} that a similar result obtains for the average energy of the pattern. On times long compared with the horizontal diffusion time, the time it takes information to diffuse across the macroscopic width $L$, the corresponding average pattern energy can be written in terms of invariant combinations of the elements of the metric and curvature tensors of the phase surface $\theta(x,y,t)$. Thus, for long times, the pattern relaxes to a configuration determined principally by the geometry of the phase surface. 

In section~\ref{sec:quarks}, we look at the consequences of a major difference between elastic and phase surfaces. In the former case, the deformed surface $z=H(x,y)$ is single valued. In the latter case, it is double valued because there is no physical distinction between labelling successive phase contours as $\theta=0,2\pi,4\pi$ or $0,-2\pi,-4\pi$. The local structure in~\eqref{local-periodic} is an even function of $\theta$. Therefore, on the plane its gradient $\mathbf{k}$ is a director rather than vector field. It is a vector field  only when lifted to a double cover of the plane. As a result, the point defects of the two surfaces are radically different. Pattern point defects generically are concave and convex disclinations with associated topological indices of $-\frac{1}{2}$ and $\frac{1}{2}$ respectively. The index measures the amount by which the director field twists as its center circumscribes the point defect in a counterclockwise direction. The twist is simply a measure of the Gaussian curvature of the phase surface which becomes concentrated at the point defect. In addition to the numerical evidence of previous work, we provide analytic/heuristic arguments as to how and why the Gaussian curvature condenses. In order to make this argument, section~\ref{sec:quarks} briefly recounts some known results upon which our conclusions depend and which also have a bearing on the structure of the additional energy which gives rise to pattern dark matter discussed in section~\ref{sec:universe}. The second part of section~\ref{sec:quarks} discusses the nature of the three dimensional analogues to concave and convex disclinations which we call loop disclinations. They bear the same relation to point disclinations as vortex loops in three dimensional contexts bear to point vortices in two dimensions. We find that now there is an additional topological invariant associated with the twist of the director field around the loop. For loop concave disclinations, this index is an integer multiple of $\frac{1}{3}$. For loop convex disclinations, it is an integer. We call these objects pattern quarks and leptons. But, as the last part of this section makes clear, there remain many outstanding challenges.

Section~\ref{sec:universe} asks the question if there are possible further parallels between patterns and phase surfaces and fundamental particle physics and cosmology. In this section, we work with a broader set of pattern forming systems which can be characterized as combinations of Hamiltonian and gradient systems. Before any phase transition has taken place, we ask how the microscopic fields evolve and make a case how the wave turbulence at these scales can produce an inflationary expansion. At the first phase transition at which a preferred scale appears, we imagine space-time to be foliated by the phase surfaces of some pattern with wavelength corresponding to the very smallest of scales. We ask about the nature of the macroscopic fields which are the slowly varying envelopes of these locally periodic structures. We show how a Goldstone mode, a mode which is neither damped nor amplified at the phase transition, can play the role of a cosmological constant. We also show that if the concentration of ordinary matter at galactic centers can cause spherical target defects to appear, there is an energy associated with the mismatch between the local pattern wavenumber and its preferred value. This energy gives rise to an additional gravitational force field which leads to galactic rotation curves entirely consistent with  Vera Rubin's observations \cite{Rubin_Rotation_1970,Rubin_Extended_1978}.

\section{Elastic sheets} \label{sec:elastic}

In a variety of condensed matter systems, multiple scale structures,
 arise spontaneously
as a result of the system attempting to find a ground state (energy
minimum) \cite{variational.review}. Somewhat crudely, these structures can be classified as {\em singularities/defects}, corresponding to small regions where energy is concentrated,  and {\em microstructure} corresponding to a (oftentimes periodic) pattern with a mesoscopic (i.e intermediate between the atomic/microscopic and the system-size/macroscopic) length scale. 
Thin elastic sheets are a prototypical example of such systems. Interestingly, depending on the configuration/forcing, they can display both singularities (d-cones, creases and their intersection points) and microstructure (wrinkling on an elastic foundation or a fluid droplet). 

We first present, a short, formal, ansatz-driven derivation of the F\"{o}ppl-von K\'arm\'an (FvK) 
energy that describes thin elastic sheets. The arguments are classical \cite{Love_Treatise_Elasticity} and we present them to set up the stage for a similar calculation for nearly periodic patterns in section~\ref{sec:patterns}. We then review rigorous results on the derivation of plate-theories for elastic sheets that are obtained by $\Gamma$--convergence~\cite{Friesecke_A_2006}, and discuss our perspective on the relation between these plate theories and the FvK functional. This perspective will also guide our discussion of the energy functionals for pattern formation in sections~\ref{sec:patterns}--\ref{sec:universe}, and thus gives a useful analogy between elastic sheets and nearly periodic patterns. The analogy also ``works" in the other direction, and we discuss  phenomena that are well known in the pattern context, eg. the zigzag and Eckhaus instabilties and their possible analogues in elastic sheets.

\subsection{3 D elasticity and asymptotic reductions for thin sheets.} \label{sec:FvKDenergy}

We will now investigate the elastic behavior of ``thin" elastic objects. Consider a two dimensional domain $\Omega$. We can model a thin sheet by the set $\Omega_h = \Omega \times [-\frac{h}{2},\frac{h}{2}] \subset \mathbb{R}^3$. The configuration of this thin sheet in $\mathbb{R}^3$ is given by a mapping $\Phi:\Omega_h \to \mathbb{R}^3$. We will use coordinates $x = (x_1,x_2,x_3)$ in the sheet and $X=(X_1,X_2,X_3)$ in the ambient space. The coordinate $x_1,x_2$ are the {\em in-plane} coordinates, and we will refer to them collectively by $x' = (x_1,x_2)$ and likewise use primes to denote in-plane differential operators, {\em e.g.} $\nabla' = (\frac{\partial}{\partial x_1},  \frac{\partial}{\partial x_2})$. $x_3$ is the coordinate in the `thin-direction', and for a sheet with constant thickness $h$, we have $-\frac{h}{2} \leq x_3 \leq \frac{h}{2}$.

\begin{figure}[htbp] 
   \centering
   \includegraphics[width=0.9 \textwidth]{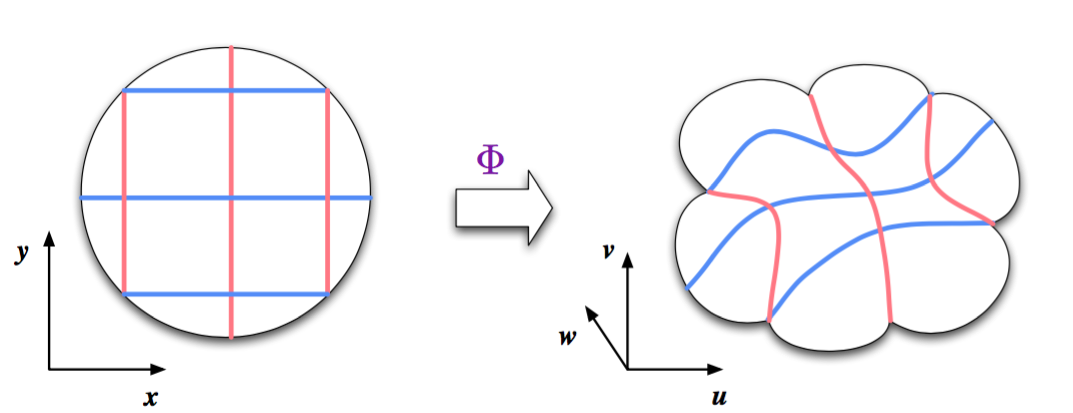} 
   \caption{Embedding a thin sheet in $\mathbb{R}^3$.}
   \label{fig:embeding}
\end{figure}

We assume that the sheet is made from a hyperelastic material, so that its stress strain relation derives from an elastic energy density $W(\mathbf{F})$, where $\mathbf{F} = D \Phi$ is the local deformation gradient. We also assume that the elastic material is isotropic. Isotropy, i.e. lack of a preferred direction in the material, and objectivity, i.e. invariance under rotations of the coordinates in the ambient space, imply that the elastic energy density satisfies $W(\mathbf{F}) = W(Q\mathbf{F}R)$ for any pair of rotation matrices $Q,R \in SO(3)$. The right Cauchy-Green tensor is defined by $C = D\Phi^T \cdot D \Phi = \mathbf{F}^T\cdot \mathbf{F}$ is also the induced metric for  the deformation. Indeed
$$
dX_1^2 + dX_2^2 + dX_3^2 = (dx_1 \ dx_2 \ dx_3) \mathbf{F}^T \mathbf{F}  \begin{pmatrix} dx_1 \\ dx_2 \\ dx_3 \end{pmatrix}.
$$
Objectivity of the elastic energy density implies that it can be expressed in terms of the  right Cauchy-Green tensor by 
$$
W(\mathbf{F}) = \rho(\mathbf{F}^T \mathbf{F})
$$
where $\rho$ satisfies  $\rho(C) = \rho(R^T C R)$ for all symmetric, positive definite, $3 \times 3$ matrix valued function $C(x_1,x_2,x_3)$ that defines a metric on $\mathbb{R}^3$ with zero Riemann curvature. Indeed the curvature has to be zero because the metric can also be expressed in the coordinates $X_1,X_2,X_3$ as $dX_1^2 + dX_2^2 + dX_3^2$. This is the compatibility condition for the right Cauchy-Green tensor \cite{Acharya_On_1999}. 

In terms of $W$, the elastic energy of a configuration $\Phi$ is given by 
\begin{equation}
\label{elasticenergy}
\mathcal{E} = \int_{-h/2}^{h/2} \int_\Omega W(D\Phi) dx' dx_3
\end{equation}
where $dx' = dx_1 dx_2$ is the in-plane area element.
We will assume that $W \geq 0$  and $W(\mathbf{F}) = 0$ at $\mathbf{F} = I$ the identity matrix. Further, $W$ is assumed $C^2$, i.e. twice continuously differentiable in a neighborhood of the identity. This allows us to consider a wide range of energy functionals; in particular we are not limiting our consideration to linear elasticity. If we linearize the energy density about the identity (a minimum), and use the assumption of isotropy, it follows that 
\begin{equation}
\label{edens}
W(\mathbf{F}) = \mu (\sigma_1^2+\sigma_2^2 + \sigma_3^2) + \frac{\lambda}{2} (\sigma_1+\sigma_2 + \sigma_3)^2 + o(\|\mathbf{F}^T \mathbf{F}-I \|^2)
\end{equation}
where $\sigma_{1,2,3}$, the  eigenvalues of the symmetric matrix $\sqrt{\mathbf{F}^T\mathbf{F}}-I$, are the {\em principal strains}. If they are small, the matrix $\frac{1}{2}(\mathbf{F}^T \mathbf{F} - I)$ represents the linearized strains due to the deformation, so that $\sigma_1+ \sigma_2 + \sigma_3 \approx \frac{1}{2}\mathrm{Tr}[\mathbf{F}^T\mathbf{F}-I], \sigma^2_1+ \sigma^2_2 + \sigma^2_3 \approx \frac{1}{4}\mathrm{Tr}[(\mathbf{F}^T\mathbf{F}-I)^2]$. The behavior of $W$ close to $\mathbf{F} = I$ (and consequently  also the behavior of $W$ for $\mathbf{F}$ near $SO(3)$) is thus characterized by two parameters $\mu$ and $\lambda$, which we can associate with the Lam\'{e} parameters describing the linear response of the material to small forces. 

Various types of reduced models for thin sheets have been derived, by many authors going back to Euler and Kirchhoff and the classical theory is reviewed in Love \cite{Love_Treatise_Elasticity}. Typically, the reduced models obtain an effective energy in terms of the (two dimensional) configuration of the center-surface $\Omega$ and one (or more) vector fields (the directors) defined on the center surface. 

For instance the Kirchhoff hypothesis posits that the fibers in the sheet perpendicular to the center surface in the reference configuration will remain straight, unstretched and perpendicular to the center surface in the deformed configuration. An improved ansatz obtains from allowing a linear stretching/compression profile for the normal fibers, whose stretching parameter is then optimized for the elastic energy \cite{Love_Treatise_Elasticity}. 

For a sufficiently smooth map $\mathbf{v}:\Omega \to \mathbb{R}^3$, representing the center surface of the deformed configuration, the normal is defined by $\mathbf{n}(x_1,x_2) =  \mathbf{v},_1 \times \mathbf{v},_2/ \|\mathbf{v},_1 \times \mathbf{v},_2\|$. In coordinates, the center surface is given by a column vector $\mathbf{v} = (v_1(x'),v_2(x'),v_3(x'))^T$ and the unit normal to this surface is given by 
$$
\mathbf{N} = \begin{vmatrix} \mathbf{i} & \frac{\partial v_1}{\partial x_1} & 
\frac{\partial v_1}{\partial x_2}\\
 \mathbf{j} &  \frac{\partial v_2}{\partial x_1} & \frac{\partial v_2}{\partial x_2} \\
 \mathbf{k} & \frac{\partial v_3}{\partial x_1} & \frac{\partial v_3}{\partial x_2} 
 \end{vmatrix}
 \qquad \mathbf{n}(x') = \frac{\mathbf{N}}{\|\mathbf{N}\|}.
 $$ 
The preceding ansatz allows us to write the map $\Phi$ as
\begin{equation}
\label{ansatz}
\Phi(x_1,x_2,x_3) = \mathbf{v}(x_1,x_2) + (x_3 + \zeta(x_1,x_2) x_3^2)  \mathbf{n}(x_1,x_2).
\end{equation}
and we can obtain an effective energy in terms of $\mathbf{v}$ by computing the deformation gradient $\mathbf{F}(x_1,x_2,x_3)$ from this ansatz and substituting the resulting expression into $E = \int W(D\Phi) dx$ and optimizing for $\zeta(x_1,x_2)$.  We get 
$$
 \mathbf{F} = \begin{bmatrix}  \mathbf{v},_1 + x_3  \mathbf{n},_1 \ \mathbf{v},_2 + x_3  \mathbf{n},_2 \ (1+ 2 \zeta x_3)\mathbf{n} \end{bmatrix} + O(x_3^2)
 $$ 
 so that
$$ 
\mathbf{F}^T \mathbf{F} - I  \approx \begin{pmatrix}  \mathbf{v},_1^T  \mathbf{v},_1-1 + 2 x_3  \mathbf{v},_1^T \mathbf{n},_1 & \mathbf{v},_1^T  \mathbf{v},_2 + x_3 (\mathbf{v},_1^T \mathbf{n},_2 +  \mathbf{v},_2^T \mathbf{n},_1) & 0 \\
 \mathbf{v},_1^T  \mathbf{v},_2 + x_3 (\mathbf{v},_1^T \mathbf{n},_2 +  \mathbf{v},_2^T \mathbf{n},_1) & \mathbf{v},_2^T  \mathbf{v},_2 -1+ 2 x_3  \mathbf{v},_2^T \mathbf{n},_2 & 0 \\
 0 & 0 & 2 \zeta x_3 \end{pmatrix} + O(x_3^2)
$$
where we have used $\mathbf{n}^T \mathbf{v},_1 = \mathbf{n}^T \mathbf{v},_2 = \mathbf{n}^T \mathbf{n},_1 = \mathbf{n}^T \mathbf{n},_2 = 0.$  The ``in-plane" components of the strain are given by ``natural geometric objects", i.e.  
$$
(\mathbf{F}^T \mathbf{F} - I)' \approx \begin{pmatrix} E & F \\ F & G \end{pmatrix} - \begin{pmatrix} 1 & 0 \\ 0 & 1 \end{pmatrix} - 2 x_3 \begin{pmatrix} L & M \\ M & N \end{pmatrix} + O(x_3^2)
$$
where $E dx_1^2 + 2 F dx_1 dx_2 + G dx_2^2$ and $L dx_1^2 + 2 M dx_1 dx_2 + N dx_2^2$ are respectively the first (metric) and the second (curvature) fundamental forms of the center surface $\mathbf{v}(x')$, since 
$$
\begin{pmatrix} E & F \\ F & G \end{pmatrix} = \begin{pmatrix} \mathbf{v},_1 \cdot \mathbf{v},_1  & \mathbf{v},_1 \cdot \mathbf{v},_2 \\ \mathbf{v},_2 \cdot \mathbf{v},_1 & \mathbf{v},_2 \cdot \mathbf{v},_2\end{pmatrix}, \quad \begin{pmatrix} L & M \\ M & N \end{pmatrix} = \begin{pmatrix} \mathbf{v},_{11} \cdot \mathbf{n}  & \mathbf{v},_{12} \cdot \mathbf{n}\\ \mathbf{v},_{12} \cdot \mathbf{n} & \mathbf{v},_{22} \cdot \mathbf{n}\end{pmatrix}.
$$
Using these expressions, along with the energy density~\eqref{edens}, we see that the total energy is given by
\begin{align}
\mathcal{E}  = & \  h \int_\Omega \big(\mu[(E-1)^2+2 F^2 + (G-1)^2] + \frac{\lambda}{2} (E+G-2)^2\big) dx'  \nonumber \\
& + \ \frac{h^3}{3} \int_\Omega \big(\mu[L^2+2 M^2 + N^2 + 4 \zeta^2] + \frac{\lambda}{2} (L+N+2 \zeta)^2\big) dx'  \nonumber \\
& + \ O(\|E-1,F,G-1\|h^3)
\label{2dElastic}
\end{align}
Note that the third term, which arises from products between the lowest order strain, i.e the in-plane strains in the center surface and $O(x_3^2)$ contributions from two derivatives of the normal (i.e the third fundamental form) are potentially of the same order as the retained terms. We will drop these terms relying on the assumption that we are in a state that is close to being unstretched (everywhere!) so $|E-1|,|F|$ and $|G-1|$ are small. Optimizing the choice of $\zeta$, we get
$$
\zeta =  -\frac{\lambda}{2 \mu + \lambda} \frac{L+N}{2} \approx -\frac{\lambda}{2 \mu + \lambda} H(z')  + o(\|\mathbf{F}^T \mathbf{F}-I \|),
$$
where the second expression is in terms of $H$, the mean curvature at $z$, and is valid if the in-plane strain is small, i.e $E \approx 1, F \approx 0, G \approx 1$. Substituting the expression for $\zeta$ in \eqref{ansatz} and \eqref{edens}, we get a deformation $\Phi$ that is slaved to the center surface $\mathbf{v}:\Omega \to \mathbb{R}^3$ by
$$
\Phi(x_1,x_2,x_3) = \mathbf{v}(x_1,x_2) + \Big(x_3 - \frac{\lambda}{2 \mu + \lambda} \frac{L+N}{2} x_3^2 \Big) \mathbf{n}(x_1,x_2), 
$$
and we also get (formally) an elastic energy for the surface given by
\begin{align}
\label{FvKDenergy}
\mathcal{E} = \mathcal{E}_s + \mathcal{E}_b = & \  h \int_\Omega \big(\mu[(E-1)^2+2 F^2 + (G-1)^2] + \frac{\lambda}{2} (E+G-2)^2\big) dx' \\
& + \ \frac{h^3}{3} \int_\Omega \big(\mu[L^2+2 M^2 + N^2] + \frac{\lambda \mu}{\lambda + 2 \mu} (L+N)^2\big) dx', \nonumber
\end{align}
where the $O(h)$ and $O(h^3)$ terms define the {\em stretching} and {\em bending energies} $\mathcal{E}_s$ and $\mathcal{E}_b$ respectively. This functional can be extended in natural ways to describe a variety of scenarios for thin sheets including the effects of boundary conditions and external loads. Consider, for example, a thin sheet clamped along its boundary $\partial \Omega$ on to a curve $x' \to \mathbf{g}(x')$ and with a normal load given by a force $\mathbf{f}(x')$ per unit area. The stable equilibria are minimizers of the functional $J$ given by
\begin{equation}
\label{forced-clamped}
J[\mathbf{v}] = \mathcal{E} - \int_\Omega \mathbf{f}(x')\cdot \mathbf{v}(x') \, dx', \qquad \mathbf{v}(x') = \mathbf{g}(x') \text{ on } \partial \Omega,
\end{equation}
where the admissible set of configurations given in terms of a map $\mathbf{v} : \Omega \to \mathbb{R}^3$ by~\eqref{ansatz} with $\mathbf{v}$ satisfying the  associated boundary condition in~\eqref{forced-clamped}.

\subsection{Plate theories through $\Gamma$--convergence} \label{sec:hierarchy}

What we have shown is that as the elastic sheet relaxes to reduce the strain energy significantly so that the minimizing configurations depend on the balance between the strain and bending energies. We have to understand, however, that~\eqref{FvKDenergy} represents only the first two terms of an asymptotic expansion (in powers of the thickness $h$) of the ``true energy functional". What one would like to be able to say is that the minimizing configuration of the truncated asymptotic expansion of the energy approximates, in a well defined sense, the true minimizer of the total energy~\eqref{elasticenergy} for the original thin but three dimensional elastic sheet. It could be that the minimizer of $\mathcal{E}_s+\mathcal{E}_b$ causes the remainder of the asymptotic expansion (formally the higher order terms will contain higher and higher derivatives) to become large compared to $\mathcal{E}_s$ and $\mathcal{E}_b$. This observation is one among significant criticisms \cite{Truesdell_Comments_1977,Ciarlet_A_1980} of the approach that leads to the {\em reduced} models~\eqref{FvKDenergy}~and~\eqref{forced-clamped} starting from full 3D elasticity~\eqref{elasticenergy}. There have been various attempts to address these criticisms and obtain rigorously justified reduced models for elastic plates by asymptotic expansions of the configuration and the loads in powers of the thickness $h$~\cite{Ciarlet_A_1980,Fox_A_1993}. However, these methods are essentially limited, as they cannot hope to describe behaviors that are outside the assumed ansatzes/asymptotic expansions for the configurations of the sheet \cite{Friesecke_A_2006}. For example, the energy functional~\eqref{forced-clamped} cannot capture the fact that, because they can {\em crumple}, thin sheets do not resist compression the same way they resist tension \cite{Steigmann141,Lecumberry_Stability_2009}.

A rigorous approach obtaining reduced dimensional models for variational problems is through the notion of $\Gamma$-convergence \cite{braides}. Given a family of energy functional $F_n:X \to [0,\infty]$, on a first countable space $X$, that depend on a parameter $n$ (which we think of as $1/h$ for elastic sheets), the $\Gamma$--limit is a functional $F : X \to [0,\infty]$ that has the following properties \cite{braides}:
\begin{enumerate}
\item For all admissible $x \in X$ and all sequences $x_n \to x$ in $X$, we have the lower bound property $F(x) \leq \lim_{n \to \infty} F_n(x_n)$.
\item For all admissible $x$, there is a sequence $x_n \to x$, the recovery sequence, such that $\lim_{n \to \infty} F_n(x_n) = F(x)$.
\end{enumerate}
The key point is that the limit functional $F$ is {\em independent of $n$}, and it describes the large $n$ behavior (leading order asymptotics) of the minimizers functionals $F_n$. The physical intuition for this procedure is that the limit functional $F$ is obtained by {\em integrating over} or {\em coarse graining} the small scale behavior, to obtain a macroscopic (i.e. $h$ independent) description of potentially small scale singular behavior in the limit $h \to 0$. The microscopic details of the small $h$ structure, e.g. crumples in a thin elastic sheet, are now encoded in the recovery sequence $x_n$ associated with the $\Gamma$--convergence $F_n \overset{\Gamma}{\to} F$.

Clearly, this procedure is useful only insofar as the limit functional $F$ is $O(1)$ (i.e. not zero or infinite) on the relevant states. This underscores the importance of appropriately scaling the energy. For thin elastic sheets, the scaling of the energy is not {\em a priori} clear. It is set by the boundary conditions \cite{Lobkovsky_Scaling_1995,MbAYP97} and the forcing, and is reflected in the nature of the defects that appear in the sheet \cite{Lobkovsky_Boundary_1996,Cerda_Conical_1999,Venkataramani_Lower_2004,Conti_Confining_2008}. The energy scale is thus an `emergent' property. The requirement that we first rescale the elastic energy to make it $O(1)$ is therefore a substantial impediment to obtaining $\Gamma$--limit energy functionals that describe the vanishing thickness limits for thin elastic sheets. 

Progress on this question requires restrictions on boundary conditions/forcing, in order that one can obtain the scaling of the elastic energy. Friesecke, James and M\"{u}ller \cite{Friesecke_A_2006} have obtained rigorous $\Gamma$--limits for the functional $J$ in~\eqref{forced-clamped} describing elastic plates with external loads. They consider  two settings in \cite{Friesecke_A_2006} -- (i) free boundaries (no clamping) and an external loading that satisfies $\int \mathbf{f}(x') dx' = 0$ (no net force) and $\int \mathbf{v}(x') \times \mathbf{f}(x') dx' = 0$ (no net moment); and (ii) fully clamped boundary conditions, i.e. $\mathbf{g}(x') = (x',0)$.  In either case, they show that for external loads that scale like $h^3$, the stretching and bending energy density (per unit area) scale as $h^5$ (``equipartition"), and the F\"oppl-von K\'arm\'an energy~\eqref{FvKDenergy} (divided by $h^5$) is the appropriate $\Gamma$--limit.

From the results in \cite{Friesecke_A_2006}, one might get the impression that the reduced energy~\eqref{FvKDenergy} is {\em incorrect} unless the energy scale is extremely small, $O(h^5)$. While this is true in the sense that  the FvK energy~\eqref{FvKDenergy} is not (a rescaling of) a $h$-independent limit energy, except in the case of very small strains and rotations (strains $\sim h^2$, curvatures $\sim h$), this is perhaps not the whole story. The energy~\eqref{FvKDenergy} is still useful, even when it is not a $\Gamma$--limit, i.e. when it is interpreted as a functional that depends on $h$, the small parameter. Indeed, the energy~\eqref{FvKDenergy} is used extensively in the physics literature (e.g. in \cite{Lobkovsky_Scaling_1995,MbAYP97,Cerda_Conical_1999,Audoly_Buckling_2008c,Mahadevan1740}) as well as for rigorous analysis (e.g. in \cite{Belgacem_Rigorous_2000,Venkataramani_Lower_2004,Conti_Confining_2008,Kohn_Analysis_2013}) even in situations where the energy {\em does not} scale like $h^5$. For blistering in thin films attached to a substrate with elastic mismatch \cite{Gioia_Delamination_1997,Vella_The_2009}, there is {\em post-facto} justification for using the elastic energy~\eqref{FvKDenergy}; the results from this reduced energy \cite{Belgacem_Rigorous_2000} agree with the results obtained from a fully 3D elastic energy  \cite{Belgacem_Energy_2002}, although the elastic energy scales like $h^2 \gg h^5$. Likewise, for elastic ridges in crumpled sheets \cite{Venkataramani_Lower_2004,Conti_Confining_2008} the results from the F\"oppl-von K\'arm\'an energy~\eqref{FvKDenergy} agree with results from fully 3D elasticity on an energy scale $h^{8/3} \gg h^5$.

\section{Phase surfaces} \label{sec:patterns}

The goal of this section is to show that the behaviors and shapes of almost periodic patterns have features in common  with those of elastic surfaces.  In particular, for pattern forming systems whose microscopic dynamics are gradient flows, the deformations of phase surfaces and elastic surfaces are governed by free energy functionals, which in certain limits, are shown to be determined by the first and second fundamental forms, the metric and curvature, of the corresponding surfaces. For elastic surfaces, this property was already known \cite{Fox_A_1993}. For the phase surface of patterns, the regularization postulated in \cite{newell1996defects} is of this form, and our results in this paper (cf. eq.~\eqref{CNreduced}) prove this property. Although the two systems have common features, there are also significant differences which we shall discuss both in this section and in the section following that focuses on the nature of the point defects.

We begin by describing briefly the macroscopic order parameters. The behavior of the system is determined by the macroscopic order parameters together with the local periodic structure linking the microscopic field with these variables. We remind the reader that planar patterns of an almost periodic nature result from a symmetry breaking transformation wherein the uniform state of the system becomes unstable (typically to stripes or hexagonal patterns; in this paper, stripes or rolls will be the dominant periodic structure)
 as some stress parameter increases above a critical threshold. The underlying system has the full symmetries of all  rotations in the plane, so the symmetry breaking instability gives rise to multiple ``stable" planforms that are related to each other by rotations. Consequently, natural patterns consist of a mosaic of almost periodic patches of stripes or rolls (or sometimes hexagons) separated by line and point defects for times long compared to the horizontal diffusion time. We take advantage of the almost periodic pattern structure to convert the non-universal microscopic equation into a universal equation for the pattern phase $\theta(x,y,t)$ whose gradient $\mathbf{k} = \nabla \theta$ varies slowly over distances of the local pattern wavelength. Far from onset, and away from the points where the field amplitude becomes small, the phase $\theta$ and its local gradient $\mathbf{k} = \nabla \theta$ are the macroscopic order parameters. The goal now is to show how to turn a non-universal microscopic energy $E$ into a universal functional $\langle E \rangle$ for the macroscopic or averaged free energy which governs the behavior of the macroscopic order parameters, or equivalently the phase surface $\theta(x,y)$.

We emphasize that the nature of the macroscopic evolution equations and its corresponding energy functional depends only on gross features of the original microscopic equation, namely that is a gradient flow, and has the symmetries of translations and rotations in the plane. There is a large set of pattern forming systems with these features. However, in order to provide a concrete example, we will work with a well known microscopic model for pattern formation, the    
Swift-Hohenberg equation \cite{SH}, which, although not a rigorous limit, is used to model high Prandtl number convection layers. That equation reads,
\begin{equation}
\label{Swift-Hohenberg}
w_t  = R w - (\nabla^2+1)^2 w - w^3  = - \frac{\delta E}{\delta w}
\end{equation}
for the real valued scalar field $w(x,y,t)$, where $R$ is the stress parameter. 

The corresponding free energy is
\begin{equation}
\label{SHenergy}
E =  \int \left[\frac{1}{2}\left( (\nabla^2+1)^2 w\right)^2 - \frac{1}{2}R w^2 + \frac{1}{4}w^4 \right]\ dxdy.
\end{equation}
For $R < 0$, the uniform state $w = 0$ is a global attractor. For $R > 0$, the uniform state destabilizes to a locally (almost) periodic roll pattern 
\begin{equation}
w = f(\theta; \{A_n\},R) = \sum A_n(k) \cos(n \theta).
\label{eq:rolls}
\end{equation} 
In~\eqref{eq:rolls}, $f(\theta)$ is a $2 \pi$ periodic function of $\theta = \mathbf{k} \cdot \mathbf{x}$ and $\mathbf{k}$ is the local wavevector. The amplitudes $\{A_n(k)\}$ are slaved to the modulus $k$ of the wavevector reflecting the rotational symmetry of \eqref{Swift-Hohenberg}. They are determined by a coupled set of nonlinear algebraic equations. For $R$ values close to the threshold value $R=0$, it can easily be shown that all even amplitudes are zero and 
\begin{equation}
\label{amplitudes}
A_1(k) \simeq R - (k^2-1)^2, \quad A_3(k) \simeq \frac{1}{128} A_1^3(k), \ldots
\end{equation}

We now turn to modulated patterns. Since $\nabla \theta$ changes slowly,~\eqref{eq:rolls} is no longer an exact solution of the original microscopic equation~\eqref{Swift-Hohenberg}. Corrections to~\eqref{eq:rolls}, involving forcing depending on gradients of $\theta$ and its derivatives must be sought. These corrections $w_1,w_2,\ldots$ arising at successive orders of powers of $\epsilon$,
\begin{equation}
\label{expansion}
w = f(\theta; \{A_n\},R) + w_1 + w_2 + \cdots
\end{equation}
satisfy the stationary equation~\eqref{Swift-Hohenberg} linearized about $w = f(\theta)$. Because the original system is translationally invariant, $w=f(\theta+\theta_0)$ is also a solution for all $\theta_0$. Therefore $f_\theta$ is a nontrivial solution of the linearized equation. The right hand sides of the equations of the successive corrections must therefore satisfy solvability conditions. These solvability conditions lead to the phase diffusion equation and related equations.

It is important to understand the anatomy of the successive corrections. At order $\epsilon$, the determination of the first correction $w_1$ gives rise to the leading order terms $-\nabla \cdot \mathbf{k} B(k)$ in the phase diffusion equation
\begin{equation}
\label{cross-newell}
\tau(k) \frac{\partial \theta}{\partial t} = - \nabla \cdot \mathbf{k}B(k) - \tau(k) \nabla^4 \theta = -\frac{\delta F}{\delta \theta}
\end{equation}
where $\tau(k) = \langle f_\theta^2 \rangle$, the average over a period of the square of the derivative of $f(\theta)$. The next correction to the phase equation occurs at order $\epsilon^3$. 

At even orders the solvability equations are automatically satisfied. However, at order $\epsilon^2$, the RHS of the equation for the correction $w_2$ will contain terms of the form $\frac{\delta^2 A_1}{A_1}$ where $\delta^2$ involves second and higher order spatial derivatives acting on $A_1$. For $A_1$ of a fixed $O(1)$ size as $\epsilon \to 0$, these terms with $A_1$ in the denominator can be incorporated in $w_2$. If $A_1$ is small, however, they cannot be incorporated into $w_2$ because the expansion~\eqref{expansion} is not uniformly asymptotic. In this instance, one must remove such terms by correcting the algebraic equations~\eqref{amplitudes}. The insight as to how to handle this difficulty came from Dave Benney in the context of making Whitham's equations compatible with the nonlinear Schrodinger equation in the small amplitude limit. This procedure turns the equation for $A_1$ into a differential equation. Indeed, no longer are the amplitudes slaved to the phase gradient, but rather, they become active order parameters. When combined with the phase equations, the amplitude correction gives a new equation for the complex order parameter $w = A e^{i \theta}$. This is precisely the Newell-Whitehead-Segel equation \cite{Newell_Finite_1969,Segel_Distant_1969}, valid near onset where the parameter $R$ in~\eqref{Swift-Hohenberg}~and~\eqref{amplitudes} is small and positive. Such a correction is also necessary near zero (complex) amplitude defects such as vortices and dislocations.

The fact that the amplitudes can become active order parameters is one of the two fundamental differences between elastic sheets and pattern phases. There is also a second major difference connected with fact that the elastic surface deformation $H(x,y)$ is single valued and the phase function $\theta(x,y)$ is double valued. We discuss this difference when we examine the nature of the point defects of the latter in section~\ref{sec:quarks}.

In this paper, we wish to derive the correction to the energy functional which leads to the biharmonic regularization term in~\eqref{cross-newell} directly. We avoid the difficulty with active amplitude parameters by working only at $R$ values far from onset and by excluding consideration of patterns with either long-lived or stationary dislocations. This approach allows us additional insights. Recall there are terms at order $h^3$ in the energy functional expansion for elastic sheets which do not conform to the FvK form of the bending energy, i.e an expression purely in terms of the curvature 2-form. We learned there that these terms become higher order once the strain energy, the order $h$ contribution, becomes small in the relaxation towards the minimum. There are similar terms at order $\epsilon^3$ when we average the microscopic energy functional. But these will also become higher order in $\epsilon$ when the pattern energy minimizes its leading $O(\epsilon)$ contribution through relaxation of the bulk wavenumber $k$ towards it preferred value $k_0$. In short, averaging the full microscopic energy functional directly allows us to see how the invariant combinations of the metric and curvature tensors of the phase surface emerge as the dominant contribution to the average energy.

Our approach, therefore, is not to iteratively obtain solvability conditions for~\eqref{Swift-Hohenberg}, but rather to directly average the microscopic energy functional~\eqref{SHenergy}. This approach is analogous to the approach Whitham \cite{whitham_book} used in his pioneering work on nonlinear wave modulations. Instead of using the nonlinear PDEs directly, he formed the modulational equations by first creating the average Lagrangian \cite{whitham_book}.
 
The gradient of $\theta$ and indeed the amplitudes $A_n(k,R)$ depend only on the long scales 
\begin{equation}
\label{scales}
 X = \epsilon x, \quad Y = \epsilon y, \quad T = \epsilon^2 t
 \end{equation}
 where $\epsilon \ll 1$ is the inverse aspect ratio, the ratio of the pattern wavelength to some macroscopic scale.  As a consequence of these scalings, the gradient operator $\nabla_{\mathbf{x}}$ is given by
 \begin{equation}
 \label{rescaling}
 \nabla_{\mathbf{x}} \to \mathbf{k}\,\partial_\theta + \epsilon \, \nabla_{\mathbf{X}}, \qquad \mathbf{X} = \epsilon \mathbf{x} = (\epsilon x, \epsilon y).
 \end{equation}
 Using this relation in \eqref{SHenergy} and \eqref{Swift-Hohenberg} respectively, we get
 \begin{equation}
 \label{rescaled-energy}
 E = \frac{1}{2} \int \left[\left( (k^2 \partial_\theta^2 +1)w+ \epsilon(2 \mathbf{k}\cdot\nabla_X + \nabla_X \cdot \mathbf{k}) w_{\theta} + \epsilon^2 \nabla_X^2 w\right)^2 - R w^2 + \frac{1}{2} w^4 \right]\ dx dy
 \end{equation}
 and
 \begin{align}
 \label{rescaledSH}
 (k^2 \partial_\theta^2 +1)^2 w -Rw =  - w^3 - w_t  
  & -  \epsilon  \left\{(2 \mathbf{k}\cdot\nabla_X + \nabla_X \cdot \mathbf{k}) (k^2 \partial_\theta^2+1) w_\theta \right. \\ 
  & \left. + (k^2 \partial_\theta^2+1) (2 \mathbf{k}\cdot\nabla_X + \nabla_X \cdot \mathbf{k}) w_{\theta} \right\}  \nonumber \\
  & +  \epsilon^2  \bigg\{(2 \mathbf{k}\cdot\nabla_X + \nabla_X \cdot \mathbf{k})^2  w_{\theta \theta} + \nabla_X^2 (k^2 \partial_\theta^2 +1) w \nonumber \\ 
  &  + (k^2 \partial_\theta^2 +1) \nabla_X^2 w \bigg\}  \nonumber 
 \end{align}
 We now average $E$ as given by \eqref{rescaled-energy} over a period of $\theta$ defining $\langle \cdot  \rangle $ as $ \frac{1}{2 \pi} \int_0^{2 \pi} \cdot \,d\theta$. The leading order term (obtained by setting $\epsilon = 0$ in \eqref{rescaled-energy})  is given by 
 \begin{equation}
 \label{eq:E0}
 \bar{E}_0 = \frac{1}{2} \int \bigg\langle k^4 w_{\theta \theta}^2 + 2 k^2 w w_{\theta \theta} + w^2 (1-R) + \frac{1}{2} w^4 \bigg\rangle dx dy.
 \end{equation} 

\begin{figure}[htbp] 
   \centering
   \includegraphics[width=0.75 \textwidth]{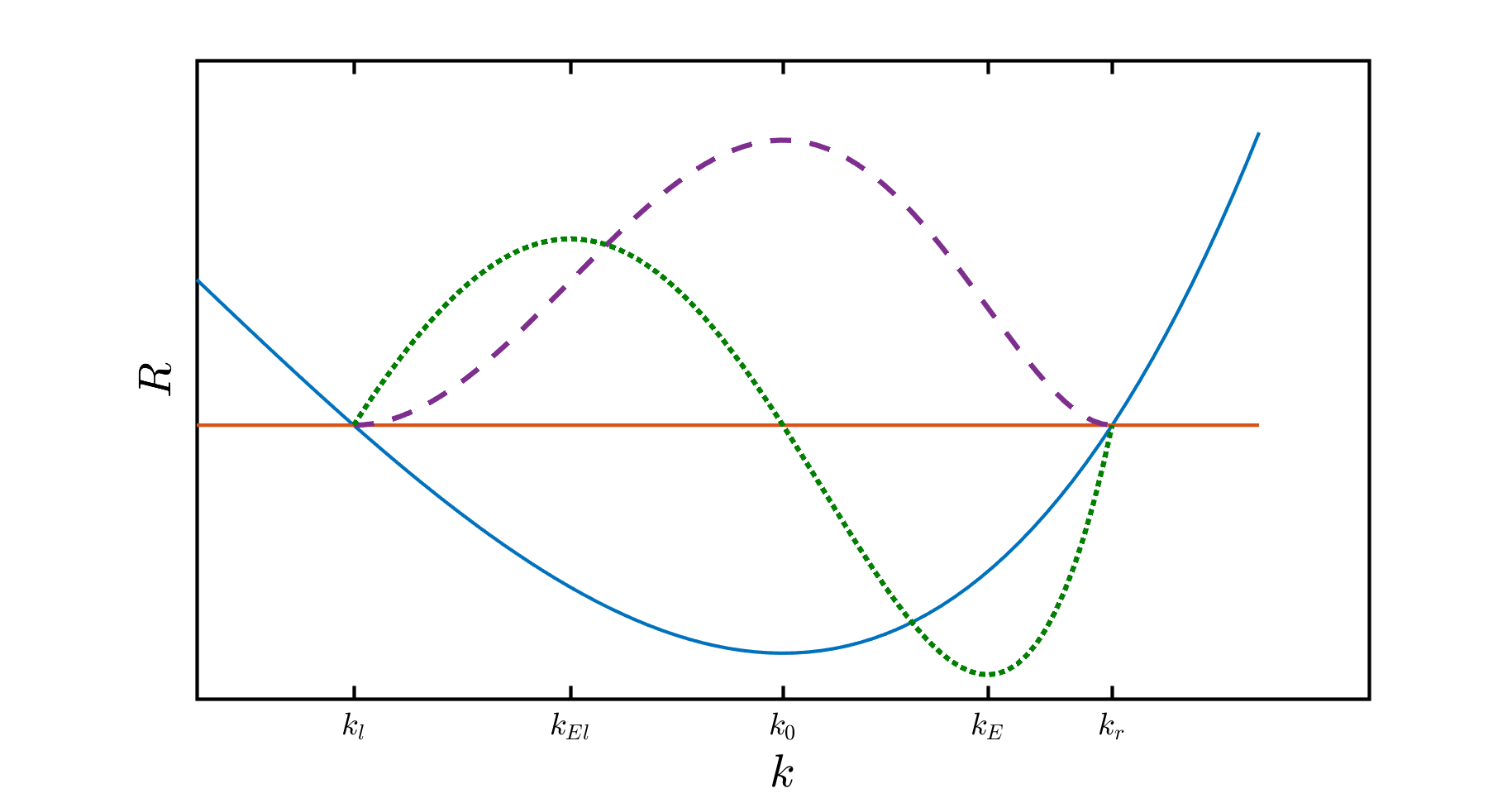} 
   \caption{The graphs of $ \langle w^4 \rangle$ (dashed) and $k B(k)$ (dotted) superposed on  the neutral stability curve (solid) of the uniform $w=0$ solution of \protect{\eqref{Swift-Hohenberg}}. The horizontal line corresponding to the given stress parameter $R$ intersects the neutral stability curve at wavenumbers $k_l$ and $k_r$. The uniform state is linearly unstable to perturbations with wavenumbers $k_l < k < k_r$. The wavenumbers corresponding to the Eckhaus instability $k_{El}$ and $k_E$, and the largest growth rate, $k_0$, are also indicated on the figure.}
   \label{fig:patterns-energy}
\end{figure}

To simplify, we integrate the first term in~\eqref{eq:E0} by parts to obtain $\int \langle k^4 w_{\theta \theta}^2 \rangle dx dy = \int \langle w k^4 \partial_\theta^4 w \rangle dx dy$. Using this identity along with \eqref{rescaledSH} with $w_t =0, \mathbf{k} $ and its derivatives frozen, in \eqref{eq:E0} we can eliminate $\partial_\theta^4 w$ to obtain
\begin{align}
\label{eq:e00} 
\overline{E}_{0} & = - \int \frac{1}{4} \langle w^4 \rangle \ dx dy 
\end{align}
To ensure convergence of the integral, we calculate $\langle w^4 \rangle$ as the difference between it value at $k^2$ and its far field value when $k^2 =  k_0^2 =1$, which is the maximum of the graph of $\langle w^4 \rangle$. To leading order therefore the averaged behavior of \eqref{Swift-Hohenberg} gives 
\begin{equation}
\label{CNflow}
\langle w_\theta^2 \rangle \theta_t = - \frac{\delta \overline{E}_0}{\delta \theta} = - \nabla_{\mathbf{x}} \cdot \mathbf{k}B(k), 
\end{equation}  
where $B(k) = \frac{1}{2} \frac{d}{dk^2} \langle w^4 \rangle$. In fig.~\ref{fig:patterns-energy}, we depict the graphs of $ \langle w^4 \rangle$ and $A$ times its derivative with respect to $k^2$ on top of the graph of the neutral stability curve.

Equation~\eqref{CNflow} is the unregularized {\em Cross-Newell} equation. It is universal for patterns forming  systems where the preferred patterns are rolls or stripes, and where the phase transition occurs between time-independent states. The graph of $\langle w^4 \rangle$ is zero at the edges $(k_l,k_r)$ of the neutral stability boundary. It rises through inflection points at $k_{El}$ and $k_E$, the Eckhaus stability boundaries, to a maximum at the preferred wavenumber $k_0$, here $k = k_0 = 1$. In general, $k_0$ will depend on the stress parameter $R$. For the Swift-Hohenberg equation~\eqref{Swift-Hohenberg} it does not. The derivative $kB(k)$ has a cubic shape. The right hand side of~\eqref{CNflow} is $-\frac{d}{dk}(kB(k)) \theta_{xx} - B \theta_{yy}$ where $x$ and $y$ are (locally) the across and along the roll directions. The Busse balloon \cite{BusseJ} region in Fig.~\ref{fig:patterns-energy} where this expression is elliptic negative (and~\eqref{CNflow} is well-posed) is $(k_0,k_E)$. For $k < k_0, B>0$, the {\em zig-zag instability} converts the mode $e^{i k x}$ into the combination $e^{i k x \pm i \sqrt{k_0^2 - k^2} y}$ with wavenumbers $k_0$ and manifests as alternate expressions of both modes separated by a line defect. In this case,~\eqref{CNflow} is regularized by~\eqref{cross-newell}. For $k  > k_E, \frac{d}{dk}(k B(k)) > 0$, the system manifests the {\em Eckhaus instability}, where rolls develop two dislocations (which require amplitude regularization) that move apart thereby removing roll pairs so as to return $k$ closer to $k_0$. 

The zig-zag and Eckhaus instabilities are generically present in pattern forming systems where the preferred patterns are rolls and the homogeneous state is unstable to perturbations with a range of wavenumbers \cite{Cross_Convection_1984}. We therefore expect they may also occur in elastic sheet in situations that lead to nearly periodic patterns, e.g. the wrinkling of an elastic thin film bonded to a compliant substrate \cite{Cerda_Geometry_2003,Audoly_Buckling_2008c,Kohn_Analysis_2013}. The variational analysis in \cite{Kohn_Analysis_2013} identifies the minimum energy states in this system with a {\em herring-bone pattern} and determine the associated wavelength. This analysis also shows that the homogeneous flat state is unstable to a range of wavenumbers. Unless they follow a particular protocol, experiments typically observe complex {\em Labyrinth patterns} \cite{Yang_Harnessing_2010} rather than a periodic herring-bone pattern \cite{Lin_Spontaneous_2007}. A natural question is whether these complex patterns are manifestations of secondary (zig-zag for example) instabilities in this system \cite{Huang_Nonlinear_2005}. We might also expect that the wrinkled films on a foundation will display the Eckhaus instability if the wrinkle wavelength is ``too small". Such a states can be achieved by a variety of mechanisms including (i) Anisotropic compression where the effective forces along the wrinkles are tensile \cite{Mahadevan1740}, (ii) Inhomogeneities in the substrate/thin film leading to a spatial variation in the local preferred wavenumber (iii) Strain softening behavior in the substrate which can decrease the preferred wavenumber with time or with increasing pattern amplitude.

Returning to the discussion of roll patterns in the Swift-Hohenberg equation, near $k =1 $, we can write the leading contribution to the average free energy $F$ as 
\begin{align}
\label{averaged-energy}
\overline{E}_0 & = \int dx dy \left( \frac{1}{4}\int_1^{k^2} \frac{d}{dk^2} \langle w^4 \rangle dk^2 \right) \\
& \approx \int dx dy \left( -\frac{1}{8} \left. \frac{d^2}{d(k^2)^2} \langle w^4 \rangle \right|_{k^2=1} (k^2-1)^2 \right) \nonumber \\
& =  -\frac{1}{4} \frac{dB}{dk^2} (k^2=1)  \int (k^2-1)^2 dx dy \nonumber
\end{align}
Since \eqref{Swift-Hohenberg} is a gradient flow for the energy \eqref{SHenergy}, the pattern will attempt to relax to rolls all having the same preferred wavenumber $k =1$.
In the vicinity of the line and point defects, higher order terms in the energy become relevant, as they regularize the singularities/discontinuities in the solutions given by the leading order ``stretching energy" which is nonconvex. We will now examine the corresponding higher order contributions to the energy for almost periodic patterns driven by gradient flows.

It is relatively straightforward to compute the averaged energy including the contributions from corrections to $w$ at order $\epsilon$ and $\epsilon^2$. The calculation reveals that $O(\epsilon)$ term in the averaged energy $E$ indeed vanishes as one expects from the symmetry $w \to - w, \epsilon \to -\epsilon$ of the system. To order $\epsilon^2$, $\bar{E}$ is given by 
\begin{align}
\label{CNregularized}
\bar{E} & =  \int \frac{1}{4}  \langle w^4 \rangle \bigg|_{k^2}^1 dx dy + \epsilon^2 \int  \bigg\{ (\nabla \cdot \mathbf{k})^2 \langle w^2_\theta \rangle + 2 (\nabla \cdot \mathbf{k}) (\mathbf{k}\cdot \nabla k^2) \frac{d \langle w^2_\theta \rangle}{dk^2}   \\
&  + (\mathbf{k}\cdot \nabla) (\mathbf{k} \cdot \nabla) \langle w^2_\theta \rangle \bigg\} dx dy .\nonumber
\end{align}
The term $\nabla \cdot \mathbf{k}$ is essentially the mean curvature of the phase surface $\theta(x,y)$. We can add an extra term $\int (\theta_{xx}\theta_{yy} - \theta_{xy}^2) dx dy $, that is related to the Gaussian curvature, because it is a {\em null Lagrangian}, i.e the area integral can be reduced to an integral on the boundary, which, in this case, measure the total twist of the wavevector $\mathbf{k}$ as one traverses the path along the boundary. In contrast to the $O(h^2)$ contribution in the normalized elastic sheet energy $h^{-1} \mathcal{E}$ from \eqref{FvKDenergy}, the $O(\epsilon^2)$ energy for nearly periodic patterns contains terms besides ones that can be reduced to expressions in the mean and Gaussian curvatures. As we shall see below, however, these additional terms become lower order when $k^2$ is almost everywhere close to its optimal value, here 1.

Let us now convert this result into the language which makes clear that the average energy depends on symmetry invariant combinations of the phase surface metric and curvature forms. The phase surface is given by the map
\begin{equation}
x' = x, y'=y, z' = \frac{1}{k_0} \theta(x,y),
\label{phase-surf}
\end{equation}
where we scale $z'$ by $1/k_0$, $k_0$ is the preferred wavenumber. In this way, when $\theta$ advances by $2 \pi$, $z'$ advances by $2 \pi/k_0$, the preferred wavelength.
In~\eqref{CNregularized}, $k_0 = 1$. The coefficients $E,F,G,L,M,N$ of the first two fundamental forms of the phase surface $z' = \theta(x',y')/k_0$ are
\begin{align} 
\label{2-forms}
E  = 1 + \frac{\theta_x^2}{k_0^2}, \quad F  = \frac{\theta_x \theta_y}{k_0^2}, \quad G & = 1 + \frac{\theta_y^2}{k_0^2}, \nonumber \\
L  = \frac{\theta_{xx}}{\sqrt{k^2+k_0^2}}, \quad M  = \frac{\theta_{xy}}{\sqrt{k^2+k_0^2}}, \quad N & = \frac{\theta_{yy}}{\sqrt{k^2+k_0^2}},
\end{align}
where $k^2 = \theta_x^2 + \theta_y^2$. If the integrands in $\bar{E}$ are functions of the traces and/or determinants of the matrices $\displaystyle{\begin{pmatrix} E & F \\ F & G \end{pmatrix}, \begin{pmatrix} L & M \\ M & N \end{pmatrix}}$, then the average free energy $\bar{E}$ will be invariant under all Euclidean motions of the $xy$ plane.

We observe from~\eqref{CNregularized} that the first term satisfies this requirement since $\displaystyle{\frac{1}{4}  \langle w^4 \rangle \big|_{k^2}^1}$ is a function only of $k^2$ which can be expressed in terms of either the trace $E+G$ or the determinant $EG-F^2$. However, the second term in the integrand is not of this form. Indeed, this terms even contains components involving third and higher order derivatives of $\theta$, {\em which clearly cannot be expressed in terms of $E,F,G,L,M$ and $N$.} Nonetheless, there is an asymptotic (in time) sense in which $\bar{E}$ only depends on (Euclidean transformation) invariant combinations of $E,F,G,L,M,N$. On the horizontal diffusion time scale proportional to $1/\epsilon^2$ (in the convection context, $d^2/\kappa$ is the diffusion time scale, $d$ the layer depth, $\kappa$ the thermal diffusivity, $\epsilon = d/L$, the inverse aspect ratio, so that $(1/\epsilon^2)(d^2/\kappa) = L^2/\kappa$, the time it takes for the pattern on one side of the box to influence the pattern on the other side), the minimization of the first term $\bar{E}$ will cause the local waveneumber $k$ to relax to its preferred value $k_0$ (here equal to 1) almost everywhere. For $k^2$ close to $k_0^2 = 1$ almost everywhere, the second and the third parts of the $O(\epsilon^2)$ component of the integrand in~\eqref{CNregularized} simplify to become lower order. 
As a result, for long times on the horizontal diffusion time scale, the flow can be expressed as 
\begin{equation}
\label{CNreduced}
\langle w^2_\theta \rangle \frac{\partial \theta}{\partial t} = - \frac{\delta}{\delta \theta} \left\{ \int \frac{1}{4} \langle w^4 \rangle \bigg|_{k^2}^1 + \epsilon^2 \langle w^2_\theta \rangle \left((\nabla_{\mathbf{x}}\cdot \mathbf{k})^2 - 2(\theta_{xx} \theta_{yy} - \theta_{xy}^2)\right) dx dy \right\}.
\end{equation}
The average energy $\bar{E}$, up to $O(\epsilon^2)$, is now a combination of $E,F,G,L,M,N$ and is invariant under translations and rotations. The steps leading to~\eqref{CNreduced}  have obvious parallels to the corresponding, ansatz-driven, reduction for elastic sheets that leads  from the energy~\eqref{elasticenergy} to~\eqref{2dElastic} and then to~\eqref{FvKDenergy}. In the elastic sheet case, in sec~\ref{sec:hierarchy}, we  make a connection between the reduced energy~\eqref{FvKDenergy} and $\Gamma$--limits for~\eqref{elasticenergy} on various scales $h^\alpha$ \cite{Friesecke_A_2006}. This brings up a natural, and as yet unresolved question:
\begin{itemize}
\item In section~\ref{sec:hierarchy} we discussed some of the potential pitfalls in relating the minimizers of the full problem to minimizers of a reduced energy obtained by asymptotic expansions. The same concerns are also present in our derivation of the averaged energy $\bar{E}$ for patterns in~\eqref{CNregularized}. Can we obtain a hierarchy of pattern energies at various orders of $\epsilon$, the inverse aspect ratio, by $\Gamma$--convergence methods applied to a microscopic equation, say~\eqref{Swift-Hohenberg}? As we discuss in section~\ref{sec:hierarchy}, we expect that the energy in~\eqref{CNreduced}, including $\epsilon$ as a small parameter, is likely applicable even on energy scales where it is {\em not} a $\Gamma$--limit.
\end{itemize}

Despite the formal similarities in the energy functional and their dependences on the two fundamental forms of the elastic/phase surface, there is a fundamental difference between the configurations that minimize the energies in the two cases. We are now about to meet the objects which are the consequences of these differences, concave and convex disclinations in two dimensions and loop concave and convex disclinations in three dimensions.

\section{Pattern quarks and leptons} \label{sec:quarks}

The principal goal of this section is to argue that the Gaussian curvature of the phase surface, which initially is spatially distributed, condenses under the action of the phase diffusion equation evolution onto point defects in two and loop defects in three dimensions. As a result, the resulting pattern contains objects which have natural topological indices usually associated with the notions of ``spin" and ``charge". Loop concave disclinations have invariant indices which are integer multiples of $\frac{1}{2}$ and  $\frac{1}{3}$ and, by analogy with particle physics, we call them pattern quarks. Loop convex disclinations have invariant indices which are integer multiples of $\frac{1}{2}$ and 1 and thus are referred to as pattern leptons. They are the analogues (and indeed the building blocks) of the point and line vortices which are observed in superconductors, in three dimensional numerical calculations conducted on the complex Ginzburg-Landau equation \cite{Aranson_Instability_1997}, and are analogous to the scroll waves observed in chemically active tissue~\cite{Keener_Scroll_1992}. In these latter contexts, the invariant indices are integer valued. For the patterns we study in this paper whose microscopic fields are real, the invariants can be fractional. What is intriguing is that such patterns, arising from microscopic equations with only translational and rotational symmetries and in which there is no ``a priori" introduction of any $U(1), SU(2)$ or $SU(3)$ symmetries, produce objects with invariant indices which are integer multiples of $\frac{1}{3},\frac{1}{2}$ and 1.

Why do they occur? Whereas energetic considerations determine the nature of the winning planforms, which in the contexts discussed in this paper are stripes or rolls in two dimensions (and planes in three), they do not choose orientation. Orientation of the roll patches results from local biases. The upshot is that natural patterns in two dimensions consist of a mosaic of patches of almost straight rolls with different directions which meld together along line defects. If neighboring patches meet on a defect line at too sharp an angle, the line defect is energetically unstable  and gives rise to concave ($V$), convex ($X$) disclination pairs, a situation described in point~\ref{VX_instability} of section~\ref{sec:point-defects}. Their structures and some realizations of their occurrence in experiments from convection to optics are shown in Fig.~\ref{disclinations-expts}. The topological character of disclinations is characterized by the winding number (twist) of the map (See fig.~\ref{fig:twist})
\begin{equation}
\mathbf{x} = (r \cos \alpha, r \sin \alpha) \to \mathbf{k} =(f,g) =(k \cos \phi, k \sin \phi)  
\label{eq:winding}
\end{equation}
the pattern order parameter, as the defect point is circumscribed. In pictorial terms, it is the angular amount $T$ through which the order parameter director field $\mathbf{k}$ turns as the position vector $\mathbf{x}$ circumscribes the singular point in a counterclockwise direction. $T(V)=-\pi$; $T(X)=\pi$ with angle average twists of -1/2 and +1/2 respectively.  Because the successive constant phase contours can be labelled either $0, 2\pi, 4\pi,\ldots$ or by $0, -2\pi, -4\pi,\ldots$ the gradient field, $\mathbf{k}= \nabla \theta$, is a double valued (director) field on the plane or a vector field on the double cover of the plane and therefore it takes two revolutions for the wave vector to return to its original value. Moreover, one cannot reconstruct $\mathbf{k}$ from a knowledge of the original microscopic field $w$ (e.g. $w=A \cos(\theta)$ has multiple solutions $\pm \theta + 2 k \pi$ corresponding to gradients $\mathbf{k} =\pm \nabla \theta$) and its gradient. Equivalently, the twist $T$ can be written
\begin{equation}
T = \frac{1}{2} \oint_C d \phi,
\label{eq:double_cover}
\end{equation}
where $C$ is a circumscribing contour on the double cover on which $\mathbf{k}$ is a vector field. The Jacobian of the map $\mathbf{x} \to \mathbf{k}$ 
\begin{align}
J & = f_x g_y - f_y g_x = \theta_{xx} \theta_{yy} - \theta_{xy}^2 , \nonumber \\
& = \frac{k}{r} (k_r \phi_\alpha - \phi_r k_\alpha),
\label{eq:gaussK_polar}
\end{align}
which is proportional to the Gaussian curvature of the phase surface $\theta(x,y)$, is a {\em very important object}. It is related to the twist by
\begin{equation}
2 J = \nabla \cdot k^2 \mathbf{P},
\label{eq:divergence_jacobian}
\end{equation}
where $\mathbf{P}$ is $(\phi_y,-\phi_x)$ (or $(\frac{1}{r} \phi_\alpha,-\phi_r)$ in polar coordinates).  If $\Omega$ is a domain on which the wave vector is smooth (in case of $V$ or $X$, $\Omega$ will be a double cover of the plane) and $C$ is a counterclockwise circumscribing contour,
\begin{equation}
2 \int_\Omega J(x,y) dxdy = \oint_C k^2 d \phi,
\label{eq:total_twist}
\end{equation}
where $C = \partial \Omega$ and encloses the singularities (if any) of $J$ in  the interior of $\Omega$. If $k = k_0$ on the boundary, then $\int J dx dy = T$.

It is easy to show from \eqref{cross-newell} (see \cite{newell1996defects}) that $J$ satisfies a conservation law, and so, from \eqref{eq:divergence_jacobian} and the conservation law, its integral \eqref{eq:total_twist} is both determined by boundary values and is conserved in time. Under the evolution~\eqref{cross-newell}, it condenses on points.

\begin{figure}[bth!]
    \centering
    \begin{subfigure}[htbp]{0.4\textwidth}
        \includegraphics[width=\textwidth]{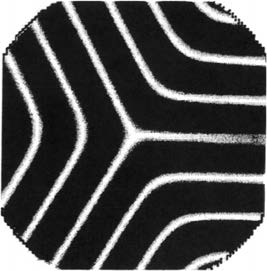}
        \caption{Concave disclination.}
        \label{concave}
    \end{subfigure}
\quad     
    \begin{subfigure}[htbp]{0.4\textwidth}
        \includegraphics[width=\textwidth]{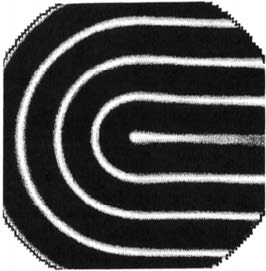}
        \caption{Convex disclination.}
        \label{convex}
    \end{subfigure}
    \caption{Point defects in patterns}
\end{figure}

\begin{figure}[h!]
  \centering
\includegraphics[width=0.75\textwidth]{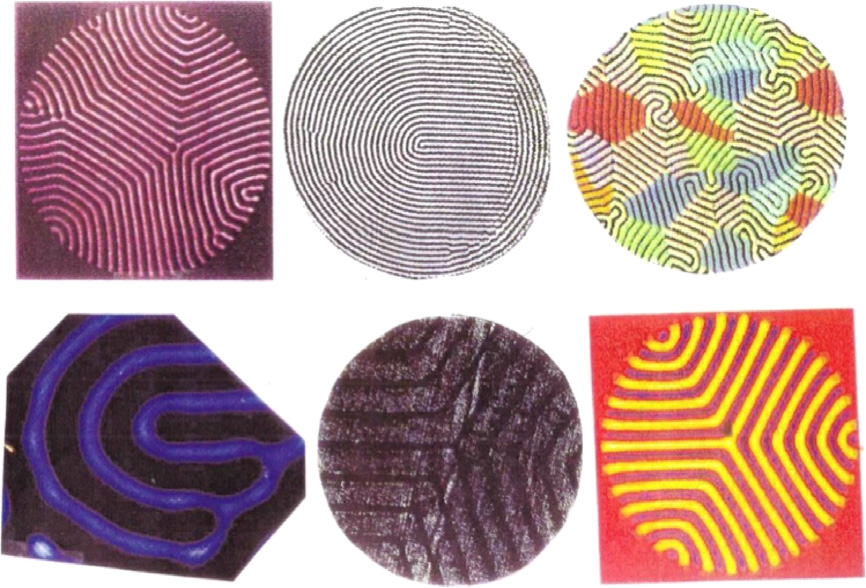}
  \caption{Concave-€"convex point disclinations in experiments.}
  \label{disclinations-expts}
\end{figure}

The indices reflect the condensation of the Gaussian curvature of the phase surface theta onto point defects in two dimensions and onto loops or line defects in three dimensions. In two dimensions, we support this conjecture with several arguments. The first is provided by a numerical experiment, first reported in \cite{newell1996defects}, in which we begin with a phase field $\theta(x,y) = \mathrm{Re}(z^3/2), z=x+iy$, on the double cover of the plane with a circular boundary. This initial field has boundary values that give the field a total twist of $-\pi$ but its Gaussian curvature is widely distributed. The phase field was evolved by the Cross-Newell equation \eqref{cross-newell} on the double cover of the plane containing the defect point \cite{newell1996defects}. Initially, the phase contours deform so as to achieve a local wavenumber $k_0$, the preferred value for $k$. This leads to the formation of phase grain boundaries or line defects along which the local wave-vector is discontinuous and the Gaussian curvature has  localized along these defect lines. As the pattern further evolves, the three defect lines become straight and intersect at 120 degrees and the Gaussian curvature condenses onto their point of intersection, a concave disclination as shown in Fig.~\ref{concave}.  Fig. 12 of \cite{newell1996defects}  clearly illustrates that the Gaussian curvature has become localized on this point defect. This condensation is also observed in recent work on disclinations in nematic liquid crystals~\cite{Zhang_Nematics_2017}, where the authors develop a novel method for the dynamics of director fields that obviates the need to work on a double cover. The other argument is heuristic and discussed in points~\ref{Gauss_condense}~and~\ref{degree} of~\ref{sec:point-defects}. Although we expect similar kinds of arguments to obtain in three dimensions, to date they have not been made. 

Compositions of point disclinations (see \cite{Passot_Towards_1994} for a discussion) such as a dislocation $VVXX$ can be seen in special circumstances but generally are not robust and disintegrate into their component disclinations. They will however dominate near pattern onset because, for very small amplitudes, the amplitude is no longer slaved to the phase gradient and becomes an independent order parameter. When combined with the phase $\theta$, it gives rise to a complex order parameter field $w=A \exp(i \theta)$ and then it and its gradient determine a unique value of $\mathbf{k}=\nabla \theta$. Thus, near onset, disclinations will recombine into composites. Away from onset, most composites will disintegrate into isolated disclinations or non-coincident disclination pairs. It is the far from onset case in which we are interested here.
To summarize: We conjecture, based on reasonable analytical and numerical evidence, that the natural point (in two dimensions) and loop (in three dimensions) defects of pattern forming systems whose pre-bifurcation uniform states enjoy translational and rotational symmetries, exhibit, without the a priori introduction of any other special symmetries, objects which one can plausibly  term quarks and leptons.
To provide a coherent picture and to explain some of the claims made in the introduction to this section, we briefly list some results reported in previous publications \cite{newell1996defects,Passot_Towards_1994,Ercolani_The_2000}.

Our starting point is the regularized Cross-Newell equation~\eqref{cross-newell} and its variational equivalent~\eqref{CNreduced}. Because the relaxation dynamics will drive the preferred wavenumber $k$ close to $k_B$, here unity,we can, after rescaling $(\theta\rightarrow\epsilon\theta, (x,y) \rightarrow (\epsilon x, \epsilon y))$ write the average energy as
\begin{equation}
\frac{1}{\epsilon} \overline{E} = \int \left[\left( 1-|\nabla\theta|^2 \right)^2 + \epsilon^2 \left(\nabla^2\theta\right)^2\right]  dxdy, 
\label{CNred}
\end{equation}
This energy is formally identical to the~{\em Aviles-Giga} energy~\cite{Aviles_Giga_1987}, although, the minimizing configurations, and the allowed defects, are very different because of $\nabla \theta$ needs to be interpreted as a director (and not a vector) field for patterns \cite{patt_defects}.

There are also connections  as well as and major differences with the Ginzburg-Landau free energy discussed by Bethuel, Brezis and Helein in \cite{GLvortices}. For the Ginzburg-Landau energy, the underlying field is a complex-valued map on a 2 dimensional domain, in contrast to~\eqref{CNred}, where we can interpret $\nabla \theta$ as a complex-valued field, but we have the additional constraint that $\nabla \theta$ is  curl free. For the Ginzburg-Landau energy, the minimizing configuration is a field of point vortices whose number and positions are determined. On the other hand, for~\eqref{CNred}, we are seeking  minimizing configurations among fields which are curl free but which also belong to a much wider class in that $\mathbf{k}= \nabla \theta$ is now a director field and not a vector field. As a result, the minimizing energies are smaller and the minimizing configurations have point defects which are disclinations rather than their vortex composites and line defects (phase grain boundaries). We now briefly list some previous results for two dimensional patterns, which provide important context for the 3 dimensional results discussed further below.

\subsection{Point defects in two dimensions} \label{sec:point-defects}

\begin{enumerate}
\item \underline{The Self-dual property:} 
Solutions of the self dual equations
\begin{equation}
\epsilon\nabla^2 \theta \simeq s(1-(\nabla\theta)^2), \qquad s=\pm 1, 
\label{self-dual}
\end{equation}
are solutions of the full fourth order Euler-Lagrange equation for~\eqref{CNred} (also steady solutions for~\eqref{cross-newell}), if the solution (level surface) $\theta(x,y)$ has zero Gaussian curvature. Even if the Gaussian curvature is nonzero, one can still minimize~\eqref{CNred} by setting $\epsilon \Delta \theta = s((1-|\nabla \theta|^2+ \chi), \theta = s \ln \psi, \chi = v \psi$ \cite{newell1996defects} to obtain the pair of equations $\Delta \psi - (1+\chi)\psi = 0, \Delta v - (1+\chi)v = - 4 \psi^{-1} J$, where $J = \theta_{xx} \theta_{yy} - \theta_{xy}^2$, as in~\eqref{eq:gaussK_polar}, is proportional to the Gaussian curvature. This was the formulation used in the numerical experiment reported in \cite{newell1996defects} showing the condensation of $J$ under evolution given by~\eqref{cross-newell}. We will use an analog of these self-dual solutions when we introduce dark matter fields in section~\ref{sec:dark-matter}.

\item \label{point_PGB} \underline{The phase grain boundary (PGB):} Straight defect line solutions turn out to have zero Gaussian curvature.  The self-dual property allows us to linearize the Euler-€"Lagrange equation via the above transformation whence $J = \chi = 0$ and~\eqref{self-dual} becomes
\begin{equation}
\epsilon^2 \nabla^2 \psi-\psi = 0 
\label{hopf-cole}
\end{equation}
Exact solutions such as $\psi = e^{s\mathbf{k}\cdot\frac{\mathbf{x}}{\epsilon}}$, $|\mathbf{k}|=1$, correspond to stripes. A linear combination of exponentials, $\psi = e^{s\mathbf{k_{+}}\cdot\frac{\mathbf{x}}{\epsilon}} + e^{s\mathbf{k_{-}}\cdot\frac{\mathbf{x}}{\epsilon}}$, $|\mathbf{k_{+}}| = |\mathbf{k_{-}}| = 1$, to a phase grain boundary (PGB) whose direction is $\frac{\mathbf{k_{+}}+\mathbf{k_{-}}}{2}$. The phase for the PGB is given by (we write the expression in terms of the original phase $\theta$ and space variables $\mathbf{x}(x,y)$)
\begin{equation}
\theta = \frac{\mathbf{k_{+}}+\mathbf{k_{-}}}{2} \cdot \mathbf{x} + \epsilon s \ln\cosh \left(s \frac{\mathbf{k_{+}}+\mathbf{k_{-}}}{2}\cdot \frac{\mathbf{x}}{\epsilon}\right) 
\label{PGB-phase}
\end{equation}
with corresponding wavevector 
\begin{align}
\nabla\theta & = (f,g) \nonumber\\
& =  \frac{\mathbf{k_{+}}+\mathbf{k_{-}}}{2} +\frac{\mathbf{k_{+}}+\mathbf{k_{-}}}{2}\tanh\left( s \frac{\mathbf{k_{+}}+\mathbf{k_{-}}}{2} \cdot \frac{\mathbf{x}}{\epsilon} \right) 
\label{PGB-k}
\end{align}
For $s \frac{\mathbf{k_{+}}+\mathbf{k_{-}}}{2} \cdot \mathbf{x} \lessgtr 0$,
\begin{equation}
\nabla \theta \rightarrow 
\begin{cases} \mathbf{k_{+}}\\ \mathbf{k_{-}}.
\end{cases} 
\label{kplusminus}
\end{equation}
The PGB has a boundary layer structure in which the wavevector $\mathbf{k}$ undergoes a transition from $\mathbf{k_{-}}$ to $\mathbf{k_{+}}$ within several wavelengths of the PGB. One can also interpret it as a weak shock solution of the hyperbolic equation $\nabla \cdot \mathbf{k} B (k) = 0$ which is regularized by the biharmonic term in~\eqref{hopf-cole} (see \cite{Ercolani_The_2000}). The solutions~\eqref{PGB-phase} and~\eqref{PGB-k} are single-valued fields $\theta$ and $\nabla \theta$ (i.e. the gradient is a vector field) that minimize the free energy~\eqref{CNred}. The PGB energy is proportional to $\sin^3\varphi$ per unit length where $\mathbf{k_{\pm}} = (\cos\varphi,\pm\sin\varphi)$. Its energy is also proportional to the square of the mean curvature $\nabla^2\theta$ of the phase surface integrated over an area containing the PGB. 

\item \label{VX_instability} \underline{VX pair creation via instability:} For large enough angles $\varphi$, the PGB solution is unstable to a director field perturbation leading to $VX$ pair creation. Why is this? Clearly if $\varphi \rightarrow \frac{\pi}{2}$, the two sets of stripes $\mathbf{k_{+}}$ and $\mathbf{k_{-}}$ become parallel to both the PGB (which now becomes a maximum of the real field $w$ and each other. The fact that
they point in different directions does not affect the cost because
as we have pointed out, it matters not at all to the real field $w$ whether the phase contours corresponding to its maxima are labeled $\theta = -2\pi, 0, 2\pi$ or $\theta = 2\pi, 0, -2\pi,...$ But according to our formula the cost, proportional to $\sin^3\varphi$, is maximum at $\varphi=\frac{\pi}{2}$. Therefore one would expect some kind of instability or phase transition to occur when $\varphi$ reaches a certain value. It does. For large $\varphi$, one can significantly lower the energy by making a transition to a director field corresponding to the creation of a $VX$ pair, as shown in fig~\ref{nipple-instability}. The principal cost in the new configuration arises again from the two PGBs
which replace a unit length of the original one. They are longer by a factor of $\frac{1}{\cos{\left( \frac{\pi}{4} + \frac{\varphi}{2} \right)}}$ but the angles they make with $\mathbf{k_{+}}$ and $\mathbf{k_{-}}$ are $\frac{\pi}{4} - \frac{\varphi}{2}$. Their combined energy is $2 \frac{\sin^3{\left( \frac{\pi}{4} - \frac{\varphi}{2} \right)}}{\cos{\left( \frac{\pi}{4} + \frac{\varphi}{2} \right)}} = 1 - \sin\varphi$.
This is lower than $\sin^3\varphi$ when $\varphi \geq 43^{\circ}$. Therefore, the $VX$ pair configuration is preferred for $\varphi > \varphi_C \simeq 43^{\circ}$. This instability and $VX$ pair creation marks a distinct departure from the analogy with elastic surfaces. There $\theta = H$, the height of the sheet above a given level, is single valued. Moreover, an elastic surface which is bent along a rooftop requires finite energy. On the other hand, as we have already stressed, a set of parallel rolls at the preferred wavelength requires no energy just because the phase contours are labeled $-4\pi, -2\pi, 0, -2\pi, -4\pi$ rather than $-4\pi, -2\pi, 0, 2\pi, 4\pi$.

\begin{figure}[h!]
  \centering
\includegraphics[width=0.6\textwidth]{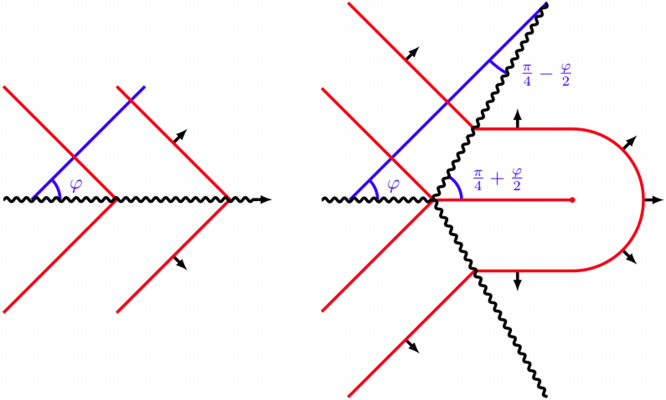}
  \caption{The nipple instability.}
  \label{nipple-instability}
\end{figure}
 
We emphasize the point. The creation of $VX$ pairs results from a phase transition/instability which occurs when two patches with different orientations (made possible by the rotational symmetry) meet at too sharp an angle. Therefore, in natural patterns where the patches have director fields corresponding to all directions, there is a high probability of the creation of concave, convex disclination pairs. This instability has been observed in experiments involving convection in elliptical containers \cite{elliptical-convection}. Note that, even saddle point singularities, the merging of two concave disclinations, for which the phase grain boundaries are at angles of $45^{\circ}$ with respect to the roll wavevectors, are unstable. Far from equilibrium, saddles will disintegrate into two concave disclinations (see Figure 12 in \cite{Passot_Towards_1994}).

\item \label{near_onset} \underline{Disclinations combine disappear near onset:} As $R$ approaches zero, the pattern onset value, at a certain $R_0(\epsilon)$, which depends on $\epsilon$, the amplitude becomes an active rather than a passive (slaved) order parameter. It combines with the phase to give a complex order parameter and consequently, disclinations, which have double valued $\mathbf{k}$ values, disappear. The reason is that the complex order parameter $w = A \exp i\theta$ and its gradient $\nabla w = \nabla A \exp i \theta + i A \exp i \theta \nabla \theta$ together determine $\nabla \theta$ uniquely. Experiments indeed show that disclinations recombine at the phase transition.

\item \underline{The energies of disclinations:} Exact multivalued solutions of the stationary hyperbolic equation $\nabla\cdot \mathbf{k}B(k) =0$ have been found using a hodograph transformation \cite{Ercolani_The_2000}. When regularized using~\eqref{CNred}, they give the $V$ (concave) and $X$ (convex) disclinations shown in figs.~\ref{concave} and \ref{convex}. The concave disclinations have three phase grain boundaries and their energy scales with their size $L$. The convex disclinations require much less energy, proportional in fact to $\ln L$, where $L$ is their size.

\item \underline{Harmonic or Laplacian disclinations:} It is a useful exercise to consider the multivalued solutions of the stationary hyperbolic equation $\nabla\cdot\mathbf{k}B(k)=0$ when we set $B(k)=1$. In this case, the constraint that $k=1$ almost everywhere is lost but the topological structure of the point defects remains the same. Let $\nabla\theta = \mathbf{k} = (f,g) = (k\cos\varphi,k\sin\varphi)$ and write $\zeta = \rho e^{i\alpha} = X +iY$ and $\theta = \text{Im} \frac{2}{3} \zeta^{3/2} = \rho^{3/2} \sin{\frac{3\alpha}{2}}$. A little analysis shows $f-ig = \rho^{1/2} \exp{(i\frac{\alpha}{2})} = ke^{-i\varphi}$. Thus $\varphi = -\frac{\alpha}{2}$. As $\alpha$ travels around the defect of $\zeta =0$, $\varphi$ twists by $-\pi$. This is the Laplace concave disclination. The Laplace convex disclination is found by setting $\theta =\text{Im} 2 \zeta^{1/2}$. The Laplace disclinations arise in the theory of quadratic differentials.

\item
\label{point_concave}
 \underline{Approximate solution for the concave disclination by six PGBs:} For $\sqrt{x^2+y^2} \gg 2\pi$  (the preferred wavelength), the concave
disclination can be well represented by six PGBs, three on one plane and three on its cover. Consider fig.~\ref{double-cover} which shows two circles split up into six sectors $S1$--$"S6$, given by angles $\frac{2(n-1)\pi}{3} < \alpha < \frac{2n\pi}{3}$, $n=1,...,6$. Using the formulas given in~\eqref{eq:rolls}~and~\eqref{amplitudes}, we can write down the corresponding phases and wavevectors.

\begin{figure}[h!]
  \centering
\includegraphics[width=0.75\textwidth]{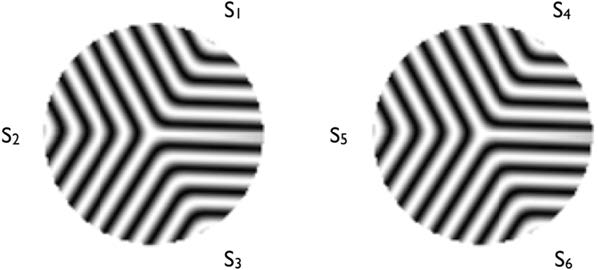}
  \caption{Sectors for a concave disclination.}
  \label{double-cover}
\end{figure}

\item
\label{point_convex}
 \underline{Approximate solution for a convex disclination:} Let $\theta = s \ln\psi$ and obtain~\eqref{hopf-cole}. On the first branch ($s = +1$), the piecewise continuous solution
\begin{align}
& \psi = e^y, x\gg 0, y>0 \quad \text{or} \quad r>0, 0<\alpha<\frac{\pi}{2} \nonumber
\\
& \psi = l_0(r) \sim \frac{e^r}{\sqrt{r}}, x<0, r>0, \frac{\pi}{2}<\alpha<\frac{3\pi}{2} \nonumber
\\
& \psi = e^{-y} x\gg 0, -y>0  \quad \text{or} \quad r>0, \frac{3\pi}{2}<0<2\pi 
\label{convex-approx}
\end{align}
can be matched in second derivative along $x = 0$ by seeking solutions for $x > 0$ in the form $e^y f(x, y)$ and neglecting the $f_{yy}$ term leading to the heat equation $f_{xx} + 2f_{y} = 0$ for $f(x, y)$. Solutions can be found which match the expansion of $\frac{e^r}{\sqrt{r}}$ for $|x| \ll y$. The second branch makes the choice $s = -1$. 

\item \label{Gauss_condense} \underline{Spin invariant and Gaussian curvature:} It is important to note that the expressions given in remarks~\ref{point_concave} and~\ref{point_convex}  are only approximate because the self dual solutions are not exact solutions of the full Euler--€"Lagrange equation for the free energy. The obstacle is a nonzero Gaussian curvature of the phase surface $\theta(x, y) $. Nevertheless, they contain the correct topologies. Any initial distribution of Gaussian curvature with twist $-ˆ'\pi$ (in the case of the concave disclination) or $+\pi$ (in the case of the convex disclination) will gather and condense on the singular point at $r = 0$. We have already discussed the numerical experiment \cite{newell1996defects} in which it was shown how a phase field with distributed Gaussian curvature, equivalently Jacobian $J= \pi k_0^2$, evolved under the action of the time dependent Cross-Newell equation and condensed first onto defect lines and then onto a single point, the center of the concave disclination. That this is not surprising can be argued heuristically. For long times on the horizontal diffusion time scale ($L^2/\kappa$, $L$ the domain size, $\kappa$ thermal conductivity; proportional to $\epsilon^{-2}$ in non-dimensional units), the first term in the energy  \eqref{CNred} must relax to a size of $O(\epsilon^2)$ in order to minimize the energy to a point where the curvature of bending terms come into play. Therefore $k^2-1$ can be at most $O(\epsilon)$ on areas. If this is so, then $J$ from \eqref{eq:gaussK_polar} has to be also small on areas. But on defect lines which evolve into straight lines, by direct calculation from point~\ref{point_PGB} above, $J=0$. But we know that the integral of $J$ over the domain is $O(1)$. It is not supported on areas or lines and therefore at best can be supported on point sets of zero measure. Since, there really is only one preferred point in the domain, the center of the concave disclination, we conclude the Jacobian $J$ and hence the Gaussian curvature condenses there. 

\item \label{degree} \underline{Topological degree, energy minimization and focusing of Gaussian curvature:} 
As we discussed in the introduction, there is strong numerical evidence that the  long time relaxation,  which is described the phase diffusion equation~\eqref{cross-newell},  concentrates the Gauss curvature $J = \mathrm{Det}(D^2\theta) = \theta_{xx} \theta_{yy} - \theta_{xy}^2$ onto point defects. This is also shown rigorously, for a situation with extra symmetry in~\cite{patt_defects}. This observation presents a mathematical puzzle since defects in minimizers of the Aviles-Giga functional concentrate energy on line defects \cite{Jin_Singular_2000}, and in the case these defects are curved, they also concentrate Gaussian curvature on lines. So why then do patterns concentrate Gaussian curvature on points? 
 
 The key difference is that $\mathbf{k}$ is a director field, that is {\em quantized} in the following sense. As we argued before, or patterns, $\mathbf{k}$ is a vector field on a double cover of $\Omega$. Considering a sequence of minimizers with $\epsilon_n \to 0$, we can use the compactness to extract a subsequential limit that will satisfy $|\mathbf{k}| = 1$ a.e \cite{Ambrosio_Line_1999,DKMO_Compacctness_2001}. Consequently, the level sets of the limit $\theta$ are (outside a ``bad" set of Lebesgue measure zero, that is potentially quite complicated), one dimensional curves orthogonal to the local direction of $\mathbf{k}$. 
 
 {\em We will assume that we can introduce branch cuts to obtain a vector field representing $\mathbf{k} = \nabla \theta$ on $\Omega \setminus C$, where the set $C$ is a union of curves that represents the branch cuts.} This assumption is physically well motivated (this is true for all the experimental and numerical patterns in this paper, figs.~\ref{convex},\ref{concave},\ref{disclinations-expts}~and~\ref{double-cover}, and is routinely used in numerical simulation \cite{newell1996defects,Zhang_Nematics_2017}), although it is not clear how one would prove this  mathematically \cite{Ambrosio_Line_1999}.  Given this assumption, a heuristic argument that illustrates why the Gaussian curvature concentrates on points. Outside of $C$, we can assume that the microscopic field $w$ is given in terms of the phase $\theta$ by~\eqref{local-periodic}, and that, for any curve $\gamma$ that intersects $C$ transversally at finitely many points, we have $\oint_\gamma \mathbf{k} \cdot d\mathbf{x} =  [\theta]_\gamma$ is the sum of the jumps in $\theta$ at the points where $\gamma$ intersects $C$. Since the branch cut set $C$ consists of curves where $\theta$ is some multiple of $\pi$, it follows that $\oint_\gamma \mathbf{k} \cdot d\mathbf{x} = m \pi$, where $m$ is an integer that depends on $\gamma$.

 For curves $\gamma$ that intersect $C$ transversally, we can define the {\em total twist} along $\gamma$ by $\oint_\gamma \theta_y d \theta_x - \theta_x d \theta_y $, which is twice the area enclosed by the image of $\gamma$ under the mapping $\nabla \theta : \Omega \to \mathbb{R}^2$. We then have,
$$
\frac{1}{2\pi} \oint_\gamma \theta_y d \theta_x - \theta_x d \theta_y = \frac{1}{2 \pi} \oint_\gamma \frac{ \theta_y d \theta_x - \theta_x d \theta_y}{|\nabla \theta|^2}  = \frac{[\theta]_\gamma}{2 \pi}
$$
is a half-integer, allowing for a possibility of twist, i.e. $\nabla \theta \to - \nabla \theta$ (cf. fig.~\ref{fig:twist}(b)), upon completing a circuit around $\gamma$. Consequently, the total twist is quantized  for curves $\gamma$ that intersect $C$ transversally. If $J$ is a signed measure such that $\int_\Gamma J dx dy$ is an integer multiple of $\frac{\pi}{2}$ for {\em every} disk inside a full (Lebesgue) measure subset of $\Omega$, it follows that $J$ is a purely atomic measure and hence is a sum of delta functions (point masses) with weights that are integral multiples of $\frac{1}{2}\pi$. 

\begin{figure}[bth!]
    \centering
    \begin{subfigure}[htbp]{0.3\textwidth}
        \includegraphics[width=\textwidth]{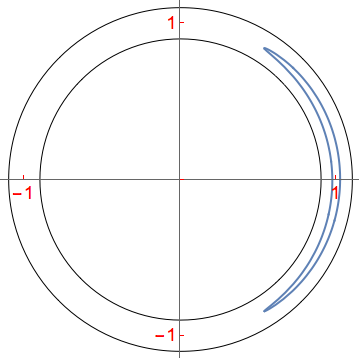}
        \caption{Winding number 0}
    \end{subfigure}
\quad     
    \begin{subfigure}[htbp]{0.3\textwidth}
        \includegraphics[width=\textwidth]{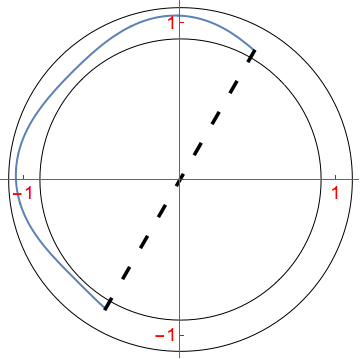}
        \caption{Winding number 1/2. }
    \end{subfigure}
\quad
\begin{subfigure}[htbp]{0.3\textwidth}
        \includegraphics[width=\textwidth]{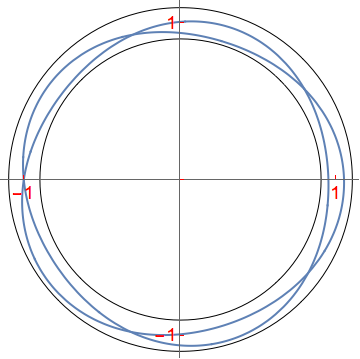}
        \caption{Winding number 2}
    \end{subfigure}    
    \caption{The figures correspond to cases where the winding numbers of $\nabla \theta(\gamma)$ are $0,1/2$ and $2$ respectively. The dashed line in (b) shows the identification $\nabla \theta \to - \nabla \theta$. The areas enclosed by the images of $\gamma$ are indeed quantized. To within $O(\delta)$, the width of the annulus, the enclosed areas are $0, \frac{\pi}{2}$ and $2 \pi$ respectively.}
    \label{fig:twist}
\end{figure}

This argument is illustrated in fig.~\ref{fig:twist} which shows the image of a closed curve $\gamma$ under a mapping $\nabla \theta$ that satisfies $||\nabla \theta| - 1| < \delta$ a.e. The area enclosed by the image of $\gamma$ is then an integer multiple of $\frac{\pi}{2}$ to within $O(\delta)$.

While this result is reminiscent of a similar result for Ginzburg-Landau vortices \cite{GLvortices} in type-II superconductors, where the magnetic flux through ``most" sets is quantized, and the magnetic field only penetrates the superconductor in point-like defects in 2D \cite{superconductivity,type_II_superconductor}, the origin of this effect for patterns is very different. Indeed the Cross-Newell energy is defined in terms of $\mathbf{k} = \nabla \theta$, and away from defects, we have the constraint $\nabla \times \mathbf{k} = 0$, like in the Aviles-Giga energy functional, and in contrast to the Ginzburg-Landau energy functional. However, if we allow for defects, then, we have $\oint \mathbf{k} \cdot d\mathbf{x} = m \pi$ is no longer necessarily zero, so that the distributional curl of $\mathbf{k}$ can be non-zero, albeit quantized, and in a coarse grained sense, it is no longer necessary that $\nabla \times \mathbf{k} = 0$. In particular, this allows for line defects in patterns to break up into point defects that carry Gaussian curvature (cf. point~\ref{VX_instability} above), while this is not allowed for single valued fields $\theta$ in the context of the Aviles-Giga functional. 

In 3D, a similar argument suggests that $J$ (now the {\em matrix} of sectional curvatures) is ``dual" to area, i.e. the integral of $J$ over small 2D disks is quantized, so that $J$ has to live on curves. We will argue below that $J$ lives on loops.

\end{enumerate}

The discussion in points~\ref{Gauss_condense}~and~\ref{degree} show that the ``spin" invariant simply reflects the condensation of Gaussian curvature on the point defect. We next argue that a new invariant, the ``charge'' is related to the condensation of the sectional Gaussian curvature on the loop backbone of the three dimensional loop concave and convex disclinations.

\subsection{Loop disclinations in three dimensions} \label{sec:loops}

Nothing in the theory outlined in Section~\ref{sec:patterns} precludes $\mathbf{k}$ from being a director field in three or more dimensions. Indeed, one can construct  three dimensional point defect analogues of concave disclinations with a tetrahedral skeleton replacing the two dimensional triangular one \cite{newell2012pattern}. Here, however, we focus on a different set of objects we call loop disclinations by attaching backbones to the two dimensional cross sections seen in figs.~\ref{concave} and \ref{convex} and asking that the real $w(x,y,z)$ field be periodic in the backbone direction $z$.

They are analogous to the scroll waves found in three dimensional autowave (excitable or oscillatory) media \cite{Aranson_Instability_1997,Keener_Scroll_1992}. Similar structures also arise as optical vortices \cite{Dennis_Local_2004,Dennis_Polarization_2008} and in nematic \cite{Zhang_Nematics_2017} and smectic \cite{Chen_Symmetry_2009,Alexander_Developed_2012,Santangelo_Curvature_2005} liquid crystals.

While in two dimensions, spiral waves (the initial pattern making instability gives rise to waves rather than stationary rolls or stripes), rotate around vortex cores, in three dimensions scroll waves rotate around filaments. But while closed (loop) filaments have twists (both along the filament and around any cross section) which are integer multiples of $2\pi$, the new objects we introduce have twist invariants which are fractional multiples of $2\pi/3$. For loop concave disclination, the twists (when divided by $2\pi$) are $\frac{1}{3}$ and $\frac{1}{2}$ respectively. For loop convex disclinations, they are $1$ and $-\frac{1}{2}$ respectively. Thus the fractional charge of $\frac{1}{3}$ appears entirely naturally. It is not, as it is in The Standard Model (TSM) of particle physics, put in by hand by a choice of the underlying governing symmetries. 

Whereas in two dimensions the hodograph transformation
linearized the unregularized phase equation $\nabla \cdot \mathbf{k}B(k)= 0$, in three dimensions we no longer have that simplification. What we do have, however, are:

\noindent (i) The three dimensional analogues of the harmonic functions $\theta = \text{Im}\frac{2}{3}\zeta^{\frac{3}{2}}$ and $\theta = \text{Im}2\zeta^{\frac{1}{2}}$, $\zeta = x+iy$ are obtained by solving
Laplace'€™s equation on $S^{1} \times \mathbb{R}^{2}$. These solutions will capture
the topology of the defects but not the constraint which chooses the length of the director field to be unity almost everywhere.

\noindent (ii) Approximate solutions to the full RCN equation
\begin{equation}
\nabla\cdot\mathbf{k}B(k) + \eta \nabla^4 \theta = 0 
\label{RCN3d}
\end{equation}
obtained by piecing together solutions of the self dual equation
\begin{equation}
\nabla^2\theta = s(1-(\nabla\theta)^2) 
\label{self-dual3d}
\end{equation}
To obtain~\eqref{self-dual3d} from~\eqref{RCN3d}, we approximate $B(k)$ by its Taylor
series about $k^2 = 1$ and rescale the spatial variables so as to
remove the factor $−\eta / k (\frac{dB}{dk^2})$ estimated at $k^2 = 1$. The self dual solutions are only approximate (they at best provide upper bounds to the associated free energy) because, in addition to the approximation for $B(k)$, they only capture the solution far from the line singularity on the backbone at $r = 0$ on which nonzero Gaussian curvature is condensed. Nevertheless, they are good in the far field and do possess the correct topologies.

Elsewhere \cite{newell2012pattern} we have given the results of these calculations. However the main ideas of the two invariants can be seen from a geometrical viewpoint. For the V string, the object of interest is a loop with a concave disclination cross-section which is twisted about the backbone so as to match the $w$ field at the two ends $z=0$ and $z=l$ which are identified. This can be done in essentially two ways. We can ask either that the phase field is periodic, i.e. $\theta(x,y,z=0) = \theta(x,y,z=l)$ or antiperiodic, i.e. $\theta(x,y,z=0) = -\theta(x,y,z=l)$. To achieve the former we must match
sectors $S_1$ and $S_3$, as shown in fig.~\ref{double-cover}, which will require a twist
of the direction $f,g,h = \nabla \theta$ along a suitable contour joining $r=r_{0}, \alpha=0, z=0$ to $r=r_{0}, \alpha=\frac{4\pi}{3}, z=l$ of $\frac{2}{3}\cdot 2\pi$. To achieve the latter, we simply match sectors of $S_1$ and $S_2$ which will require an angular twist of $\frac{1}{3}\cdot 2\pi$. Each of their negatives is also possible by twisting in the clockwise direction. We call this index the ``charge" invariant. The spin invariant is obtained by examining the twist of the direction $\nabla \theta$ around any cross-section. For the $X$-string, the field $w$ can only be made periodic in the backbone direction by twisting the backbone by an integer multiple of $2\pi$. Therefore its charge invariant is an integer multiple of $\pm 1$. The spin index again is $\frac{1}{2}$ Each invariant is associated with the twist of the director field around the two independent directions on the torus and the line integrals can be related to the area integrals of the appropriate sectional Gaussian curvature of the three dimensional hypersurface, which, we conjecture, will condense onto the loop backbone.

\begin{figure}[h!]
    \centering
    \begin{subfigure}[b]{0.32\textwidth}
        \includegraphics[width=\textwidth]{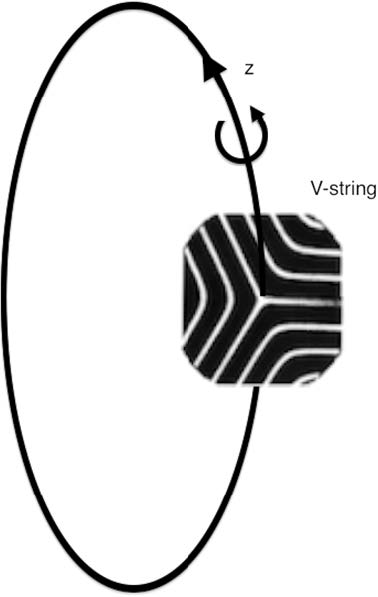}
        \caption{A V-string.}
    \end{subfigure}
    ~ 
    \begin{subfigure}[b]{0.3\textwidth}
        \includegraphics[width=\textwidth]{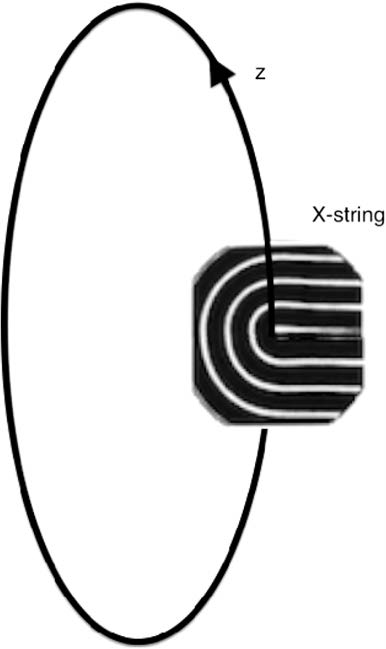}
        \caption{A X-string.}
    \end{subfigure}
    \caption{Loop defects. $z$ is the coordinate along the ``backbone" and the pattern is periodic in $z$.}
    \label{loop-defects}
\end{figure}

There is a connection of the ``charge" invariants with the Gaussian curvature of the twisting phase surface that has a boundary which consists of $C_{1}$: a helical curve joining $(r=r_{0}, \alpha = 0, z=0)$ to $(r=r_{0}, \alpha=\frac{4\pi}{3}, z=l)$ : $C_{2}$: the straight line at $\alpha=\frac{4\pi}{3}$, joining  $r=r_{0}$ to $r=0$ : $C_{3}$ : the backbone on which $k\rightarrow 0$ joining $z=l$ to $z=0$ at $r=0$ : $C_{4}$ : The straight line joining $r=0,\alpha=0,z=0$ to $r=r_{0},\alpha=0,z=0$. The value of $\frac{1}{2\pi}\int_{c_1}k^2 d\varphi$ is $\frac{2}{3}$. Its value on $C_{2}$ and $C_{4}$ is zero because on these straight lines $\left[ \varphi \right] = 0$. The value along the backbone is also zero. Thus $\frac{1}{2\pi r_0} \cdot 2 \int (\nabla f \times \nabla g)$ onto $z=0$, $r\leq r_{0}$, $0\leq\alpha\leq\frac{4\pi}{3}$ and onto $\alpha=0, 0\leq r \leq r_{0}, 0 \leq z \leq l$ is $\frac{2}{3}$.One can calculate these integrals for the case where we approximate $\theta$ by $\frac{2}{3}r^{\frac{3}{2}} \sin{\left( \frac{3\alpha}{2} - \frac{2\pi z}{l}\right)}$. Then $\frac{1}{r} (f_{r} g_{\alpha} - f_{\alpha}g_{r}) = - \frac{1}{4r}, f_{r}g_{z} -f_{z}g_{r} = \frac{\pi}{l}$ and $\frac{1}{r}(f_{\alpha}g_{z} - f_{z}g_{\alpha}) = 0$. The integral $\frac{2}{2\pi r_{0}} \int_{0}^{r_0} \int_{0}^{\frac{4\pi}{3}} \frac{1}{r} (f_{r} g_{\alpha} - f_{\alpha}g_{r}) rdrd\alpha = - \frac{1}{3}$ whereas $\frac{2}{2\pi r_{0}} \int_{0}^{r_0} \int_{0}^{l} \frac{\pi}{l} drdz = 1$. In the harmonic case, one must divide out by the radius $r_{0}$ as
the wavenumber $k^2$  does not tend to the preferred wavenuber 1 in the far field. 

Therefore the two sets of invariants, the ``spins" $-\frac{1}{2}$ and $\frac{1}{2}$ and the ``charges" $\pm 1, \pm \frac{2}{3}, \pm \frac{1}{3}$ 
reflect the amounts of sectional Gaussian curvatures which have condensed on the loop backbones. On the other hand, the energy of the $V$ string is proportional to the mean curvature condensed along the PGBs for the $V$-string which is proportional to $3\sin^3\frac{\pi}{6}$ times multiplied by the product of its cross sectional and backbone lengths $L$ and $l$. The $X$-string energy is proportional to $l\ln L$. 

The realization that energy/mass/ mean curvature concentrates on defects (points and lines in 2D and loops and planes in 3D) means that matter creation does not occur uniformly throughout space but condenses in localized regions. Therefore in a pattern universe, the clumping of visible matter is not only a consequence of self gravitational interactions but primarily follows the spatial distribution of defects. It is also remarkable that, just as the effects of mass and matter are felt through geometrical deformations, so too pattern charge and spin are realized by the condensation of quantized sectional Gaussian curvature of the phase surface on defects as well.

However, there remain many open questions and challenges.
\begin{enumerate}
\item  While we have made a plausible case for loop concave and convex disclinations, we have not yet established, either by analytical or computational means, their existence as solutions of the class of pattern forming equations we have been studying. Some encouragement is provided by the fact, however, that similar objects have been observed/computed in singular optics \cite{Dennis_Local_2004,Dennis_Polarization_2008}, liquid crystals \cite{Zhang_Nematics_2017,Chen_Symmetry_2009,Alexander_Developed_2012,Santangelo_Curvature_2005}  and chiral nematic colloids \cite{Tkalec_Knots_2011}. While in two dimensions it is easy to prescribe boundary data to make a concave disclination pattern, it is not so easy to imagine in $\mathbb{R}^3$ what far field conditions would lead to loop concave disclinations. On the other hand, if the three dimensional space consisted of one periodic direction, e.g. $S^1 \times \mathbb{R}^2$, then the loop defects shown in fig.~\ref{loop-defects} are not at all implausible.
 
 A related question is to find the three dimensional analogue of the nipple instability, discussed in point 3 of this section. For a planar geometry, two constant wavevector patches meeting along a line defect at to sharp an angle produces a $VX$ pair. What three dimensional configurations can lead to loop dislocation clusters?

\item  It would be desirable also to establish the amount of free energies associated with these objects and various pairings such as two loop concave disclinations separated by a distance, say $\mathbf{b}$. One might argue that the gradient of that free energy with respect to $\mathbf{b}$ would determine the force laws between such pairs. Perhaps, the methods in~\cite{Zhang_Nematics_2017} for defects in nematic liquid crystals can also be applied in this context.

\item  If they do exist, how do they form composites? Can they form linked loops or will linked loops invariably reconnect to form larger loops \cite{Berry_Topological_2007,Berry_Reconnection_2012}?

\item  If two, say loop concave disclinations, reconnect, is the new along the filament invariant, which we have termed the ``charge", the sum of the ``charges" of the reconnecting loops? If this is so, can one then show that the stable composites have the charges of $\pm1$ or zero? Such objects could then be thought of as pattern protons and antiprotons and pattern neutrons.

\item  If, instead of stationary lines of constant phase, the pattern consists of traveling waves, can one still have loop concave and convex disclinations? These would be the disclination analogues of scroll waves whose cores are vortices. However, because of radiation losses, in many cases, closed filaments of scroll waves collapse. In that case, are there ranges for the parameters in the governing pattern equation for which loop disclinations survive for long times?

\end{enumerate}

Attempts to answer these and related questions are ongoing and will be reported on elsewhere.

\section{Pattern universes} \label{sec:universe}

Since patterns can exhibit objects analogous to quarks and leptons, it is natural to ask if there are further pattern analogues and parallels with fundamental physics and cosmology. We are not the first to ask this question, and there is significant work in the physics literature on interpreting quantum mechanics and gravitation as coarse graining a pattern space-time/world crystal \cite{Hooft_Equivalence_1988,Kleinert_book} that is discrete/periodic on the Planck scale (See also the popular exposition \cite{witten1996reflections} for a discussion of related ideas from the vantage point of String theory). In this brief section, using simple pattern models, we explore this possibility and suggest that indeed further parallels do exist. We illustrate that, in addition to pattern quarks and leptons, we can find analogues to inflation, and how that process leads to an isotropic far field, to the notions that some information can indeed be communicated at speeds faster than the speed of light, that different physical processes can dominate at different scales, that it is natural that dominant physical constants at one scale are related to those at another, and that there are potential analogues to the ideas of, and roles played by, so called dark matter and dark energy or quintessence. 

Before we lay out a prima facie case that such an exploration may lead to interesting parallels, and is worthy of further investigation, we want to emphasize that, for the moment, what we suggest is very preliminary and in no way intended to be a new theory for the universe we live in and an explanation for how it came into being. Rather we see our contributions in much the same way that the imaginative and pioneering work of Yves Couder \cite{Couder_Single_2006, Protiere_Particle_2006}, with the help and support of many colleagues and notably John Bush \cite{Bush_Pilot_2015}, explores the possibility of using the classical physics of a particle bouncing on the surface of a sea of Faraday waves to provide parallels with quantum theory, illustrating the wave-particle synergy and duality and even reproducing the results of the two slit experiment, long thought to be the province of quantum theory alone.  What Couder has done is illustrate that many of the phenomena we think of as being purely quantum effects have classical analogues. It is in this spirit that we play with and explore what we call a {\em pattern universe}, i.e. the consequences of having additional fields in the universe that obey ``pattern-forming" equations. 

\subsection{Dark matter} \label{sec:dark-matter}

\begin{figure}[htbp] 
   \centering
   \includegraphics[width=0.6 \textwidth]{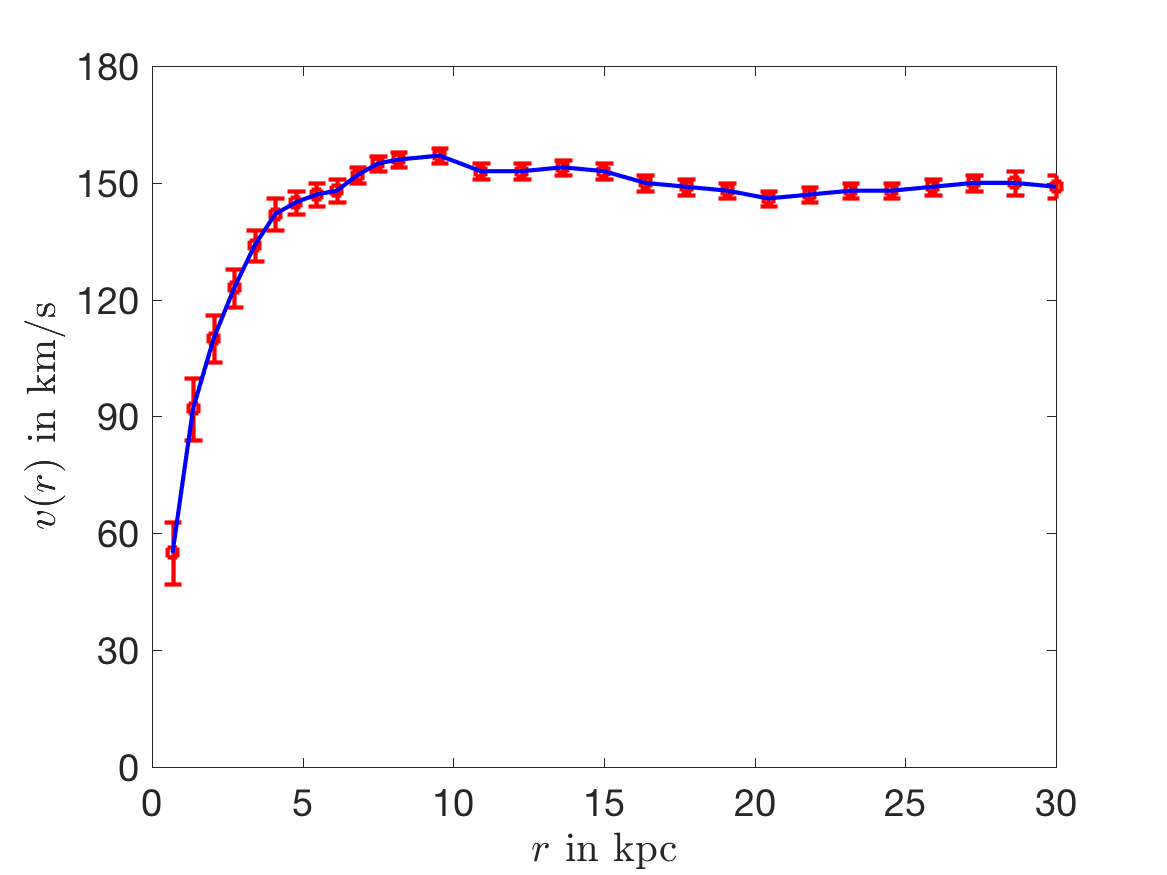} 
   \caption{The rotation curve for NGC 3198. $r$ is the distance from the galactic center and $v(r)$ is the rotation speed. The data is from van Albada {\em et al} \protect{\cite{van_Albada_1985}}.}
   \label{fig:rotation}
\end{figure}

We begin by offering the idea that a pattern universe has an additional energy which could give rise to effects consistent with those which have led to the suggestion from cosmologists \cite{Blumenthal_Formation_1984,Davis_Evolution_1985} that, in addition to visible matter, there exists a significant and additional source of gravitation whose mass is many times that of visible matter \cite{Zwicky_Masses_1937,Rubin_Rotational_1980}. Important findings which led to a widespread acceptance of this conclusion are the studies of Rubin, Ford and Thonnard~\cite{Rubin_Rotation_1970,Rubin_Extended_1978} on the rotational speeds of stars in galaxies.  The simple model in which the force experienced by the star executing a circular orbit at radius $r$ about the galactic center due to some large center mass $M$ of the galaxy is balanced by the star's centrifugal force leads to a conclusion that the rotational velocity $v$ should decay with increasing distance $r$. This is contradicted by observations that consistently demonstrate that the galactic rotation curves flatten out and the orbital velocity of distant stars is roughly constant. This is illustrated, for example, in fig~\ref{fig:rotation}  generated from rotation velocity data for the galaxy NGC 3198 (van Albada {\em et al} \cite{van_Albada_1985}).

Turning the argument around by balancing $GM/r^2$ with $v^2/r$ with $v$ constant, gives a mass $M$ of $v^2 r/G$ which is more mass than the galaxy would seem to contain. Dark matter was invented to resolve this discrepancy \cite{Zwicky_Masses_1937,Rubin_Rotational_1980}. For example, the data in fig.~\ref{fig:rotation} can be explained by a spherical distribution of dark matter in the galactic halo \cite{van_Albada_1985}. Unfortunately, to date, the source of this extra mass has not yet been definitively identified. Also, there are countervailing view points. Mordechai Milgrom \cite{Milgrom_MOND_1983} and others have argued that the observations should be interpreted as the need to modify Newton's third law at extremely small accelerations; others, most notably Don Saari, have argued that the simple balance in terms of a net mass $M$ is not appropriate and that one must take account of forces between individual stars, and this may negate the need for adding any other matter.

We consider a field $w$ that corresponds to locally periodic stripe patterns with a preferred  wavenumber $k_0$. While an accurate analysis of such a field requires us knowledge of an appropriate Lagrangian in  curved space times, expressed in terms of covariant derivatives, our initial approach is to study this problem in a background (flat) Minkowski space. Writing $w = f(\theta)$ for a periodic function $f$ of a phase variable $\theta$, the appropriate Lorentz-invariant generalization of the pattern Lagrangian~\eqref{CNred} to this setting can be obtained from the `minimal coupling' assumption \cite{MTW} as  
\begin{equation}
\bar{E} = \frac{\rho_0 c^2}{k_0^4}   \int \left\{(|\nabla \theta|^2-c^{-2}\theta_t^2-k_0^2)^2 +  (\Delta \theta-c^{-2} \theta_{tt} )^2\right\} \ d^3x \,dt.
\label{Lagrangian}
\end{equation}
where $\rho_0 c^2$ is the effective ``energy density" of the phase field, $k_0$ is its preferred wave-number is an inverse length. 
Although we do not have a concrete identification of the nature of the field $w$ nor the mechanism which leads to the local periodic structure $w = f(\theta), \nabla \theta = \mathbf{k}$, nor what sets the preferred wave-number $k_0$, we observe that~\eqref{Lagrangian} gives the {\em natural} Lorentz-invariant generalization of the {\em universal averaged energy} for nearly periodic stripe patterns~\eqref{CNred}, and is thus expected to describe the macroscopic behavior arising from a variety of microscopic models. The integral is set up with extra normalizing factors of $k_0$ so that $\bar{E}$ has the dimensions of action, i.e. energy integrated in time. The resulting Euler-Lagrange equation is a 4th-order nonlinear wave equation on a nearly-flat 3+1 spacetime. We seek spherical target solutions which both reflect the galactic halo and are ``localized", so that wavenumber mismatch $|\nabla \theta| - k_0$ vanishes as $r \to \infty$. Such solutions can be found with the self dual approximation given by
\begin{equation}
(\Delta \theta-c^{-2} \theta_{tt}) = \pm(|\nabla \theta|^2-c^{-2}\theta_t^2-k_0^2)
\label{self-dual-instanton}
\end{equation}
We can linearize this equation by a Cole-Hopf transformation $\theta = \mp \log \psi$ to obtain
\begin{equation}
\frac{1}{c^2}\psi_{tt} - \Delta \psi + k_0^2 \psi = 0
\label{schrodinger-wave}
\end{equation}
Stationary, rotationally invariant, self-dual solutions $\psi(\mathbf{x},t) = \psi(r)$ are given by
\begin{equation}
\label{yukawa}
\frac{1}{r^2} (r^2 \psi'(r))' = k_0^2 \psi(r) \quad \Rightarrow \quad \psi(r) = \frac{A e^{k_0 r} + B e^{-k_0 r}}{k_0 r}.
\end{equation}
We can make this solution smooth at the origin by picking $B = -A$ so that 
$$
\theta = - \log(\psi) = \theta_0 - \log(\sinh(k_0 r)) + \log(k_0 r).
$$
Of course, a more accurate analysis will involve solving the full fourth-order Euler-Lagrange equation for $\theta(r)$, and we expect to get corrections to the self-dual solution from the `curvature' of the phase field near $r=0$. The far field behavior of $\theta$ is no longer exactly given by $A=B$, but nonetheless, we expect that it is still given by~\eqref{yukawa}, and for $r \gg  k_0^{-1}$, the growing exponential dominates so we obtain
\begin{equation}
\theta \approx \begin{cases} -k_0 r + \log (k_0 r) + \theta_0 + \log(2) + O(e^{-k_0r})& r \to \infty \\ \theta_0 + \theta_2 (k_0 r)^2 + O(r^4) & r \to 0 \end{cases}
\label{full-solution}
\end{equation} 
Consequently, the energy density in the (stationary) phase field is given by 
\begin{align}
\rho c^2 &= \frac{\rho_0 c^2}{k_0^4} ( (\Delta \theta)^2 + (|\nabla \theta|^2 -k_0^2 )^2) \nonumber \\
& \sim \begin{cases} \rho_0 c^2 k_0^{-4} (2  k_0 r^{-1} - r^{-2})^2 & r \gtrsim  k_0^{-1} \\ \rho_0 c^2 O(1) & r \lesssim k_0^{-1} \end{cases}
\label{density}
\end{align} 
By the principle of equivalence, this energy density will act as source for the curvature of space-time \cite{MTW}. Assuming that the space-time is close to being flat, we can compute the galactic rotation curves in the Newtonian approximation. The gravitational acceleration due to the density distribution $\rho$ given in~\eqref{density} can be computed through its mass function $M(r) = 4 \pi \int \rho(r) r^2 dr$. Balancing this acceleration with the centripetal acceleration $v^2/r$ yields
$$
v = \sqrt{\frac{G M(r)}{r}} \sim \begin{cases}  \sqrt{4 \pi G \rho_0} r & r \lesssim  k_0^{-1} \\ \sqrt{4 \pi G \rho_0} k_0^{-1} & r \gtrsim  k_0^{-1} \end{cases}
$$
which is consistent with the observed the rotation curves, i.e. a linear profile for small $r$ followed by a flattening of the curve beyond a certain radius. To match observations (cf. fig.~\ref{fig:rotation}) the core radius $k_0^{-1}$ should be comparable to the scale-length of the galaxy.

We might therefore ask: Could dark matter simply be a manifestation of the elastic energy of a patterned structure which foliates the universe whose energy, reflecting on the deviation of the local wavenumber from its preferred value, depends only on what we see as detectable (visible and black hole)  matter? 

We have no way of answering that question definitively. There is no doubt that, in the absence of any idea about the microscopic description of the early universe, these suggestions are purely conjectural and lack a means of verification. However, we have introduced a new paradigm and the new picture has the virtue that it is potentially consistent with Rubin's observations \cite{Rubin_Extended_1978}. Also, it avoids some of the small-scale inconsistencies between observations and  the existing cosmological constant/cold dark matter ($\Lambda$CDM) paradigm \cite{Navarro_Core_2000,Weinberg_CDM_2015}, since there is a natural ``core" radius $ k_0^{-1}$ in our model

\subsection{Inflation, entanglement and dark energy in a pattern universe} \label{sec:quantum-patterns}

In this section we investigate phenomena that arise on macroscopic scales as the parameters in a microscopic ``quantum pattern equation"  are varied. The key idea is that, starting with a microscopic description, a master equation which captures behaviors at all scales, the physics at successively larger scales is revealed, stage by stage, by averaging over the dominant scales at the previous level.  Fields which vary on very long times and distances can behave very differently to fields which have very short scale variations. Among the consequences is that there need be no contradiction between super-luminal  spreading of the influences of short scale fluctuations, while the communication of ``information", a coarse-grained, large scale notion is limited by the speed of light \cite{Hooft_Equivalence_1988}. 

We will begin with the microscopic equation, closely related to the one we used in the previous section.
\begin{align}
\label{microscopic} 
\rho w_{tt}+\mu w_t = & \nabla^2(D \nabla^4+P \nabla^2+a)w +b((\nabla w)^2+2w \nabla^2 w) \\
& -4c_1w^3+2c_2(w(\nabla w)^2+w^2\nabla^2w). \nonumber
\end{align}
We have applied an extra Laplacian operator to the RHS of~\eqref{microscopic} as we want to include the effects of a Goldstone or soft mode which will turn out to parallel the Einstein cosmological ``constant" or quintessence. The stress parameter $P$ here will essentially parallel the inverse temperature in cosmological models. Alternatively, we can take $P$ fixed and let the stress parameter be $a$.  Nonlinear effects are included in the quadratic and cubic terms. For simplicity, we choose the RHS of~\eqref{microscopic} to be the variational gradient of a functional $E$, the energy,
\begin{equation}
\label{energy-univ}
E = \frac{1}{2}\int\left\{ D (\nabla(\nabla^2 w))^2 - P(\nabla^2 w)^2 + (a+bw)(\nabla w)^2 + 
c_1 w^4 + c_2 w^2 (\nabla w)^2\right\} d\mathbf{x},
\end{equation}
although, for most pattern behaviors, that restriction is not really necessary. Let us also think of the friction $\mu$ as being very small and, at least to start with, we take it to be zero.

For small $P$, and for small amplitude values of the field $w$, the motion is wavelike, $w \sim \exp(i\mathbf{k\cdot x}-i \omega t)$,with dispersion relation, 
\begin{equation}
\label{dispersion}
\omega^2=k^2\left(D\left(k^2-\frac{P}{2D}\right)^2+a-\frac{P^2}{4D}\right).
\end{equation}

Thus, at the beginning, the $w=0$ state is neutrally stable and small perturbations behave as a sea of waves. A field of weakly nonlinear, dispersive waves will interact, share energy via resonances, and that energy will more and more be spread to the smallest scales. This is basically a fundamental result of wave turbulence theory (see references \cite{zakharov2012kolmogorov,newell2011wave}) and is consistent with the idea that the system can more effectively explore its phase space (increase entropy) by transferring energy to smaller and smaller scales. It is also known from wave turbulence theory that the smallest scales exhibit isotropy and homogeneity. The speed at which the spectral energy reaches the smallest scales depends of the competition between the relative strengths of the nonlinear terms and linear terms at large $k$. For situations with finite capacity such as fully developed hydrodynamic turbulence, the smallest scales are reached in finite time. For what are called infinite capacity situations, the time can be exponential. At small wavelengths, the energy spectrum is also isotropic. Because wave-packet speeds increase with increasing $k$, the size of an initial bubble of waves will rapidly expand either at an exponential or supra-exponential rate. An initial bubble containing a turbulent field of random waves spreads at inflationary rates and the small scale structures which dominate the outermost parts of this universe are isotropic.

As the size gets larger and larger, and the stress parameter $P$ grows, there comes a stage when an instability or phase transition occurs and certain shapes and scales are preferentially amplified. If we examine the linear stability of the $w=0$ solution by setting $w \sim \exp(i\mathbf{k \cdot x}+\sigma t)$, we find, analogous to~\eqref{dispersion}, that
\begin{equation}
\label{growthrate2} 
\sigma^2=-k^2\left(D\left(k^2-\frac{P}{2D}\right)^2+a-\frac{P^2}{4D}\right).
\end{equation}
The growth rate first becomes zero when 
\begin{equation}
\label{critcal} 
 P=P_c=2\sqrt{aD}, \quad k^2 =  k_c^2=\left(\frac{a}{D}\right)^{1/2}. 
\end{equation}

As $P$ increases, the maximum amplification occurs for $P=P_c(1+s)>P_c$  at scales $k_0=k_c(1+s)^{1/4}$. Out of the random sea of, by now very weak nonlinear waves, a dominant periodic or quasi-periodic structure emerges. Rotational symmetry means of course that there are many competitors, namely all modes with wave-vectors with wavenumber $k_c$ and indeed their immediate neighbors. These modes compete for dominance via the nonlinear interactions and, eventually, a winner, or a set of possible winners depending on available symmetries, emerges. Often that winner can be a single mode, a pattern of stripes or rolls, with a wave-vector whose direction is arbitrary (and picked by local biases; i.e., symmetry breaking) but it can also be a combination of modes all with the same preferred wavenumber. Here, for simplicity, we choose the mode to be a travelling wave.  

However, for conservative systems such as~\eqref{microscopic} in the absence of friction, there are potentially other players which are not amplified but which are not damped either. The most important of these is the zero or constant mode. It represents a symmetry of the system in which an arbitrary real number is added to $w$  (analogous to the situation in a convecting fluid at small Prandtl numbers where the symmetry in the pressure field-the pressure only appears as a gradient-can drive large scale mean drift flows) is a Goldstone or soft mode and will play a very important role in the system's dynamics and in its choice of a final state.  In principle, we cannot neglect all the oscillatory modes either but we will argue here that once the first phase transition is reached, a friction which is zero at $k=0$ comes into play, and these other modes are damped. We will therefore seek to understand the evolution of the envelope $A$ of the most excited wave under self interactions and interactions with the Goldstone mode $B$. 

We insert the shape
\begin{equation}
\label{testshape} 
w=A e^{i \mathbf{k_0 \cdot x}}+\mathrm{cc}+B +\mathrm{corrections}, \quad \mathbf{k_0}^2=\frac{P}{2D} =\sqrt{\frac{a}{D}},
\end{equation}
into~\eqref{microscopic} and allow, in order to account for the fact that wave-vectors nearby to the preferred mode also play a role, both the envelope $A$ and the mean $B$ be slowly varying functions of space and time. Standard analysis \cite{newell1974envelope} leads to two equations for the evolution of both the envelope $A$ and the mean $B$. They take the form of two nonlinear wave equations,
\begin{align}
\frac{\partial^2A}{\partial t^2}-4a\nabla^2 A & = -2 b k_0^2AB + \text{growth and saturation terms} \label{eqA} \\
\frac{\partial^2 B}{\partial t^2}-a\nabla^2 B& = 2bk_0^2AA^*+ \text{cubic nonlinearities} \label{eqB}
\end{align}
The reason one does not get the familiar and canonical complex Ginzburg-Landau equation \cite{newell1974envelope} is that the frequency $\omega$ (alternatively  the growth rate $\sigma$) has a double zero at the onset value $k_0^2=\frac{P}{2D}=\sqrt{\frac{a}{D}}$. In fact, properly taking the limit, the two wave speeds, $2\sqrt{a}$ and $\sqrt{a}$, are the respective (critical) group velocities of the modes.

The equations~\eqref{eqA}~and~\eqref{eqB} tell us how all information depending on the envelope of the amplified mode $A$ and the dynamically important but neutral mode $B$ propagate. We think of $c=2\sqrt(a)$ as the ``speed of light". We make two important points.
First, while information associated with functionals of the original microscopic field $w$ can potentially travel at any speed (there is therefore nothing to contradict the phenomenon of ``entanglement"), information which is a functional only of the envelope $A$ which varies on long scales travels with a definite speed $c$ determined by the microscopic parameters, in this case is $\sqrt{2}a$. 

Second, we regard the mean field $B$ as being driven by the dc component proportional to $AA^*$ arising from the quadratic term. It is not, in any point by point manner, proportional to $AA^*$ but rather captures the integrated history of the envelope intensity. The equation~\eqref{eqA} then parallels the Einstein equation (in the near flat universe approximation) which describes how the curvature of space-time is deformed by the presence of distributed mass. The term $AB$, again arising from the quadratic interaction in~\eqref{microscopic}, plays exactly the role that the cosmological constant term would play in orthodox theory. Whereas the function $B$ is not constant, it reflects some average over the state of the intensity of the amplified fluctuations. In a pattern universe, therefore, a Goldstone mode plays the role of a cosmological constant accelerating the expansion of the universe.

In summary, then, our picture is that, before the first phase transition, the field consists of a bubble of rapidly expanding nonlinear waves. At the first phase transition, the microscopic field $w$ develops a periodic structure. Even though all information about the behavior of the envelope $A$ is contained in~\eqref{microscopic}, its behavior is dominated by the pattern averaged equations and the physics associated with long scales. Subsequent phase transitions may introduce further separations of behaviors, each associated with an envelope of the field of the most excited mode at that transition. There is no reason that the scales of the later transitions should not lie between the ``Planck" scale and the very long scales on which the equation for the envelope $A$ obtains. 

Now what happens as $P$, the stress, continues to increase (the temperature continues to decrease)? We know from previous analyses on pattern formation \cite{Passot_Towards_1994} that, while near onset the point defects have to have surrounding vector rather than director (spinor) fields, far from onset the amplitude of $A$ is determined by the gradients of its phase and that phase can be double valued. In other words, near onset the order parameter $A$ is complex. Both the amplitude and its phase (its phase gradient is the vector $k$) are active order parameters, each obeying a dynamics governed by the solvability equation~\eqref{eqA}. For ease of visualization, let us consider the case of two space dimensions. In that case, the only point defects are composites of the concave and convex disclinations discussed in section~\ref{sec:quarks}. They are saddles, vortices, targets, dislocations. However, as $P$ increases away from its critical onset value, the amplitude of the envelope becomes algebraically slaved to its phase gradient (and, because of rotational symmetry, on the modulus $k$ of the wave-vector $\mathbf{k}$). There is then only one real order parameter and the direction of the wave-vector $\mathbf{k}$ cannot be determined from the original microscopic field $w$ and its gradient. This point was previously emphasized in section~\ref{sec:quarks}. It is at this stage that pattern quarks and pattern leptons can appear. They are unbound. Indeed, composites are generically unstable as was shown clearly in \cite{Passot_Towards_1994}.  At the next phase transition, however, when the relevant order parameter again becomes complex, and therefore the phase gradient $\mathbf{k}$ is a vector, pattern quarks and leptons begin to merge to form bound states. In  \cite{Passot_Towards_1994}  we showed that, when we begin with a field containing isolated concave and convex disclinations, and then lower the stress $P$ back towards it critical value $P_c$, the isolated disclinations disappear to be replaced by composites. This next transition might then be considered as the analogue of the separation of the grand unified theory into the strong and electroweak forces. An even later transition would parallel the separation of the weak and electromagnetic forces.
 
 In summary, we have made a prima facie case that pattern or crystal universes can exhibit many features which have analogues to that vast array of behaviors observed in fundamental particle physics and cosmology, most of which, dark matter, dark energy, are as yet unexplained by orthodox theories. Our basic idea has been that the pattern universe is one containing structures at many scales, each set being determined by the preferred configurations arising at a set of discrete phase transitions. We have shown that objects with some of the same invariants as quarks and leptons arise naturally without any imposition of symmetries that might induce particular fractional values. We have shown that a universe foliated by phase surfaces can give rise to additional gravitational like forces which lead to star rotation speeds consistent with Rubin's observations. We have explained how there can be a rapid expansion of the early pattern universe leading to an isotropic far field. We have suggested that the nature of the field which promotes an accelerated universe is related to the presence of Goldstone mode released at a phase transitions. In short, we have offered a new paradigm the ultimate value of which will only be known after  further work.


\section*{Acknowledgments}
This work was supported in part by the NSF through the award DMS 1308862. We are grateful to Amit Acharya, Randy Kamien and Yves Pomeau for many stimulating discussions. We are also grateful to the two referees for a careful reading of the manuscript and their thoughtful comments, which significantly improved the paper.

\bibliographystyle{sapm}      
\bibliography{patterns-sheets}   

\end{document}